\documentclass[10pt,aps,prb,twocolumn,superscriptaddress]{revtex4-2}
\usepackage{amsmath,amssymb}
\usepackage[dvipdfmx]{graphicx}
\usepackage{bm}
\usepackage{multirow}
\usepackage{hyperref}

\usepackage{wasysym}
\usepackage{graphicx}
\usepackage{dcolumn}
\usepackage{physics}
\usepackage{diagbox}
\usepackage{ulem}

\newcommand{\Eq}[1]{Eq.~(\ref{#1})}
\newcommand{\Sec}[1]{Sec.~\ref{#1}}
\newcommand{\Fig}[1]{Fig.~\ref{#1}}
\newcommand{\App}[1]{Appendix~\ref{#1}}

\begin{document}
\preprint{APS/123-QED}

\title{Topological phase transitions by time-dependent electromagnetic fields\\in frustrated magnets: Role of dynamical and static magnetic fields
}

\author{Tatsuya Shirato}
\email{tshirato@phys.sci.hokudai.ac.jp}
\affiliation{Graduate School of Science, Hokkaido University, Sapporo 060-0810, Japan}

\author{Ryota Yambe}
\affiliation{Department of Applied Physics, The University of Tokyo, Tokyo 113-8656, Japan}

\author{Satoru Hayami}
\email{hayami@phys.sci.hokudai.ac.jp}
\affiliation{Graduate School of Science, Hokkaido University, Sapporo 060-0810, Japan}

\date{\today}
\begin{abstract}
We theoretically investigate the effects of time-dependent electromagnetic fields on frustrated magnets with the spatial inversion symmetry. 
Two types of external-field setups are considered: One is a circularly polarized electromagnetic field 
and the other is a combination of a circularly polarized electric field and a static magnetic field.
The system is modeled by a classical frustrated Heisenberg model on a triangular lattice, whose ground state is a single-$Q$ spiral spin configuration. 
The effects of irradiated electric and magnetic fields are taken into account by the inverse Dzyaloshinskii--Moriya (DM) interaction and the Zeeman coupling, respectively, without heating effects.
By numerically solving the Landau--Lifshitz--Gilbert equation,
we find that the two field configurations lead to distinct skyrmion crystal (SkX) phases and their associated topological phase transitions: in the former setup, SkXs composed of skyrmions with skyrmion numbers of one and two with opposite signs emerge, whereas in the latter setup, SkXs with the same sign appear.
The stabilization mechanisms of these SkXs are accounted for by the competition among electromagnetic-field-induced chiral DM interactions, electric-field-induced three-spin interactions, and the Zeeman coupling based on the high-frequency expansion within the Floquet formalism.
Furthermore, for the latter setup, we find that the stability region of the SkX phase varies significantly depending on the timing of the application of the circularly polarized electric field and the static magnetic field.
Our findings would broaden the possible routes to generate and control SkXs by time-dependent electromagnetic fields, advancing both the theoretical comprehension and experimental control of topological spin crystals.
\end{abstract}
\maketitle

\section{INTRODUCTION} \label{sec:introduction}
The concept of topology has attracted tremendous interest due to its fundamental significance in physics.
The study of topology in condensed matter systems may be broadly classified into two categories: topology in real space and that in reciprocal space.
The pioneering studies of the former can be traced back to the discovery of the Kosterlitz-Thouless transition~\cite{KT1973JPhysC,Kosterlitz1974JPhysC}, whereas those of the latter originate
from the foundation of Thouless-Kohmoto-Nightingale-den Nijs formula~\cite{TKNN1982PRL,KOHMOTO1985Ann.Phys}.
In the context of the real-space topology, skyrmions, characterized by an integer topological number called a skyrmion number, have been studied both theoretically and experimentally~\cite{skyrme1962unified, Bogdanov89, Bogdanov1994, Robler2006, Nagaosa2013,Zhang2020skyrmion,Tokura2021,Hayami2021itinerant,Hayami2024MaterToday}.
The skyrmion number corresponds to how many times the classical spins consisting of the skyrmion spin texture wrap a three-dimensional unit sphere.
In magnetic materials, skyrmions often emerge as periodic alignments, the so-called skyrmion crystals (SkXs), composed of superpositions of multiple spiral spin density waves with symmetry-related ordering wave vectors, i.e., the multiple-$Q$ state.
Their topological nature gives rise to a fictitious electromagnetic field via the Berry phase mechanism~\cite{Berry1984}, which generates exotic quantum transport phenomena~\cite{Loss_PhysRevB.45.13544, Ye_PhysRevLett.83.3737, Ohgushi2000, Shindou_PhysRevLett.87.116801, Neubauer2009,Yu2012}.
Magnetic skyrmions are promising candidates for building highly integrated and low-power neuromorphic devices that specialize in running artificial neural network models~\cite{Yokouchi2022Sciadv, Lee2024NatMater}.  
Because of their nanoscale size and low-power operation compared to conventional spin textures, they are expected to serve effectively as physical reservoirs in reservoir computing systems~\cite{Tanaka2019Neural}.

SkXs have been discovered in various magnets without spatial inversion symmetry since the first experimental report in chiral magnets $\mathrm{MnSi}$~\cite{Muhlbauer2009, Neubauer2009, Adams_PhysRevLett.107.217206} and Fe$_{1-x}$Co$_x$Si~\cite{yu2010real, adams2010skyrmion}.
They are stabilized by the competition between the ferromagnetic exchange interaction and the Dzyaloshinskii--Moriya (DM) interaction~\cite{Robler2006, Yi_PhysRevB.80.054416}, the latter of which arises from the relativistic spin--orbit coupling~\cite{Dzyaloshinsky1958, Moriya1960}. 
Meanwhile, the formation of SkXs has also been demonstrated in centrosymmetric magnets, where the first experimental demonstration has been done in $\mathrm{Gd_2PdSi_3}$~\cite{Kurumaji2019} and $\mathrm{Gd_3Ru_4Al_{12}}$~\cite{hirschberger2019skyrmion, Hirschberger_10.1088/1367-2630/abdef9}.
In such systems, the geometrically frustrated interaction or the Ruderman--Kittel--Kasuya--Yosida (RKKY) interaction~\cite{RK1954PhysRev, Kasuya1956PTP, Yosida1957PhysRev} leads to competing exchange interactions among spins, which can be a source of multiple-$Q$ states.  
Upon applying external stimuli such as magnetic fields or thermal fluctuations, these frustrated systems can undergo phase transitions into SkX states, where the external effects play a crucial role in their formation~\cite{Okubo2012, Koshibae2014, Leonov2015, SZLin2016PhysRevB_GL, Hayami2016bubble, Hayami2021inplane}.
Most of the previous theoretical studies have focused on the role of the Zeeman coupling under static magnetic fields~\cite{Bogdanov1989, Bogdanov1994, Robler2006}.
In recent years, increasing attention has been paid to the realization and control of SkXs through electric fields~\cite{PhysRevResearch.6.013228}. 
In particular, in certain multiferroic materials, electric fields couple directly to spins~\cite{Tokura2014}, giving rise to various phenomena such as the creation of skyrmions by electric-field pulses~\cite{Mochizuki2015, Mochizuki2016, Huang2018} and the stabilization of SkX phases under circularly or linearly polarized electric fields~\cite{Yambe2024circular, Shirato_JPSJ.94.063601_bimeron}.

In this paper, we theoretically study the effects of time-dependent electromagnetic fields on frustrated magnets.
Specifically, we consider two setups on external fields: (i) a circularly polarized electromagnetic field and (ii) a combination of a circularly polarized electric field and a static magnetic field. 
We irradiate these time-dependent electromagnetic fields with a classical frustrated Heisenberg model on a triangular lattice, whose ground state is a single-$Q$ spiral state.
The electric field is assumed to couple to the spin degree of freedom via the inverse DM mechanism~\cite{Sato2016PRL, Yambe2023}.
By solving the Landau--Lifshitz--Gilbert (LLG) equation under the applied electromagnetic fields without including heating effects, qualitatively different results are obtained for the two setups.
Under circularly polarized electromagnetic irradiation, SkXs composed of skyrmions with the skyrmion numbers of one and two but with opposite signs are generated. 
In contrast, when both a circularly polarized electric field and a static magnetic field are simultaneously applied, SkXs with the same sign skyrmion number are formed. 
The stabilization mechanisms of the obtained SkXs in each case can be elucidated in terms of electromagnetic-field-induced chiral DM interactions, electric-field-induced three-spin interactions, and the Zeeman coupling to the magnetic fields within the high-frequency expansion in the Floquet formalism.
Furthermore, for the second setup, our results reveal that the stability window of the SkX phase strongly depends on when the circularly polarized electric field and the static magnetic field are applied.

The rest of this paper is organized as follows.
In \Sec{sec:model}, we introduce the classical Heisenberg model with frustrated exchange interactions and extend it to include the effects of time-dependent electromagnetic fields.
Then, we outline the method based on the LLG equation in \Sec{sec:method}.
In \Sec{sec:results}, we show the simulation results for the right circularly polarized (RCP) electric field.
Finally, \Sec{sec:summary} is devoted to the summary. 
The simulation results for the left circularly polarized (LCP) electric field are shown in \App{sec:appendixA}.

\section{MODEL} \label{sec:model}
We consider a classical Heisenberg model with frustrated exchange interactions on the two-dimensional triangular lattice, which is given by
\begin{align}
  \mathcal{H}_{\mathrm{0}} = -J_1\sum_{\langle i,j \rangle}\boldsymbol{S}_i\cdot\boldsymbol{S}_j
                    -J_3\sum_{\langle\langle i,j \rangle\rangle}\boldsymbol{S}_i\cdot\boldsymbol{S}_j\label{H_0},
\end{align}       
where $\bm{S}_i=(S^x_i, S^y_i, S^z_i)$ represents the localized spin at site $i$ with $|\bm{S}_i|=1$.
The Hamiltonian consists of the ferromagnetic exchange interaction between the nearest-neighbor spins $\langle i,j \rangle$, $J_1>0$, and the antiferromagnetic exchange interaction between the third-neighbor spins $\langle\langle i,j \rangle\rangle$, $J_3<0$. 
We set the lattice constant to unity $a=1$. For $J_1/|J_3|<4$, the ground state is the single-$Q$ spiral state with the ordering wave vector $|\bm{Q}^*|=Q^*=2\cos^{-1}[(1+\sqrt{1-2J_1/J_3})/4]$.
There are six equivalent ordering wave vectors;
$\pm\boldsymbol{Q}^*_1=\pm Q^*(1,0)$, $\pm\boldsymbol{Q}^*_2=\pm Q^*(-1/2,\sqrt{3}/2)$, and $\pm\boldsymbol{Q}^*_3=\pm Q^*(-1/2,-\sqrt{3}/2)$.

We apply external electromagnetic fields on the model in Eq.~(\ref{H_0}).
The model Hamiltonian is given by
\begin{align}
  \mathcal{H}(t) = \mathcal{H}_{\mathrm{0}} -\bm{E}(t)\cdot\sum_{\langle i,j \rangle}\boldsymbol{p}_{ij} -\bm{B}(t)\cdot\sum_i\boldsymbol{S}_{i}, \label{H}
\end{align}
The second term in \Eq{H} represents the coupling of the electric field $\bm{E}(t)$ to the electric dipole $\bm{p}_{ij}$ between the nearest-neighbor bond;
a circularly polarized electric field is given by $\bm{E}(t)=E_0(\delta\cos\omega t, -\sin\omega t, 0)$ with the strength $E_0$, the frequency $\omega$, and the periodicity $T=2\pi/\omega$.
$\delta=+1~(-1)$ stands for the right (left) circular polarization.
The electric dipole is activated by the inverse DM mechanism in the form of $\bm{p}_{ij} = - \lambda \bm{e}_{ij}\times(\bm{S}_i\times\bm{S}_j)$ with the magnetoelectric coupling constant $\lambda$ and the unit bond vector $\bm{e}_{ij}$~\cite{Katsura2005,Mostovoy2006,Sergienko2006,Tokura2014}; 
we ignore other symmetry-allowed couplings between the electric field and electric dipoles~\cite{Yambe2023}.
The third term in \Eq{H} describes the Zeeman coupling to the magnetic field $\bm{B}(t)$. 
We consider two types of magnetic fields: One is the circularly polarized magnetic field $(-B_0\sin\omega t, -\delta B_0\cos\omega t, 0)$ with the strength $B_0$ and the other is the static magnetic field $(0, 0, B_{\mathrm{s}}^z)$ with the strength $B_{\mathrm{s}}^z$, namely, $\bm{B}(t)=(-B_0\sin\omega t, -\delta B_0\cos\omega t, B_{\mathrm{s}}^z)$.

\section{METHOD} \label{sec:method}
To obtain non-equilibrium steady states (NESSs), which is defined as a state exhibiting temporal fluctuations around a steady time-averaged value, in \Eq{H}, we solve the LLG equation by ignoring heating effects.
The LLG equation is given by
\begin{align}
  \frac{d\bm{S}_i}{dt}=&-\frac{\gamma}{1+\alpha_{\mathrm {G}}^2}\left\{ \bm{S}_i\times\bm{B}^{\mathrm {eff}}_i(t)+ \alpha_{\mathrm {G}}\,\bm{S}_i\times[\bm{S}_i\times\bm{B}^{\mathrm {eff}}_i(t)] \right\},
  \label{LLG}
\end{align}
with the gyromagnetic ratio $\gamma$, the Gilbert damping constant $\alpha_{\mathrm {G}}$, and the effective magnetic field $\bm{B}^{\mathrm {eff}}_i(t)=-\partial\,\mathcal{H}(t)/\partial \bm{S}_i$.

While solving the LLG equation, we calculate the local scalar spin chirality, which is given by 
$  \chi_{\bm{r}}(t) = \bm{S}_i(t)\cdot[ \bm{S}_j(t)\times \bm{S}_k(t)]$,
where sites $i$, $j$, and $k$ form the triangle at the position vector $\bm{r}$ in counterclockwise order.
The skyrmion number for the whole system on a discrete lattice is given by
\begin{align}
  N_{\mathrm{sk}}(t)= \frac{1}{4\pi} \sum_{\bm{r}} \Theta_{\bm{r}}(t), \label{Nsk}
\end{align}
with a skyrmion density $\Theta_{\bm{r}}(t)\in [-2\pi,2\pi)$~\cite{Berg1981};
\begin{align}
  \tan\frac{\Theta_{\bm{r}}(t)}{2} = \frac{2\chi_{\bm{r}}(t)}{[\bm{S}_i(t)+\bm{S}_j(t)+\bm{S}_k(t)]^2-1}.
\end{align}
The skyrmion number in the magnetic unit cell (the number of magnetic unit cells: $N_{\mathrm{unit}}$) is defined as $n_{\mathrm{sk}} = N_{\mathrm{sk}}/N_{\mathrm{unit}}$.
In addition, we calculate the magnetization, which is given by
\begin{align}
M^\alpha(t) = \frac{1}{N}\sum_{i}S^\alpha_i(t) \label{M},
\end{align} 
where $N$ is the system size and $\alpha= x,y,z$.
We also calculate the momentum decomposition of spins in reciprocal space, which is given by
\begin{align}
  S_{\bm{q}}(t) = \frac{1}{N}\sqrt{  \sum_{\alpha,i,j}S^\alpha_i(t)S^\alpha_j(t) e^{i\bm{q}\cdot(\bm{R}_i-\bm{R}_j)}},
\end{align}
where $\bm{q}$ is the wave vector and $\bm{R}_i$ is the position vector at the site $i$.
The averages of these quantities are calculated once the system attains a NESS.
By setting the time to reach the NESS as $t_0$, the averaged quantities are given by
\begin{align}
  O &= \frac{1}{N_{\mathrm{smp}}}\sum_{n=1}^{N_{\mathrm{smp}}} O(t_0+n\Delta)  \label{average},
\end{align}
with the number of samples $N_{\mathrm{smp}}$ and the time step $\Delta$.
The typical values of $t_0$, $N_{\mathrm{smp}}$, and $\Delta$ are set to $t_0=800T$, $N_{\mathrm{smp}}=20000$, and $\Delta=0.02T$, respectively.

\section{RESULTS} \label{sec:results}
In the following simulations, we choose $J_1=1$ and $J_3=-0.5$ with $Q^*=2\pi/5$; the ground-state energy per site is given by $E_{\mathrm{gs}}=-J_1[ \cos Q^* +2\cos (Q^*/2) ]-J_3[ \cos (2Q^*) +2\cos Q^* ]$, whose absolute value is taken as the energy-scale unit.
We set $\gamma=1$ and $\alpha_{\mathrm{G}}=0.05$.
The LLG equation is solved by using the open software DifferentialEquations.jl~\cite{rackauckas2017differentialequations}.
The time scale is set as $|E_{\mathrm{gs}}|^{-1}$.
We consider the system consisting of $N=20^2$ spins under the periodic boundary conditions; we have confirmed that qualitatively similar results have been obtained for larger system sizes, e.g., $N=40^2$. 
In the following, we focus on the high-frequency regime, where the frequency $\omega$ is large enough compared to the static energy scale $|E_{\rm gs}|$;
we fix $\omega/|E_{\rm gs}| = 7.5$.
Hereafter, we exhibit results for the RCP electric field.
Results for the LCP electric field are shown in \App{sec:appendixA}.

\subsection{Circularly polarized electromagnetic fields} \label{ssec:dynamicB}

\begin{figure}[t!]
  \begin{center}
  \includegraphics[width=1.0\hsize]{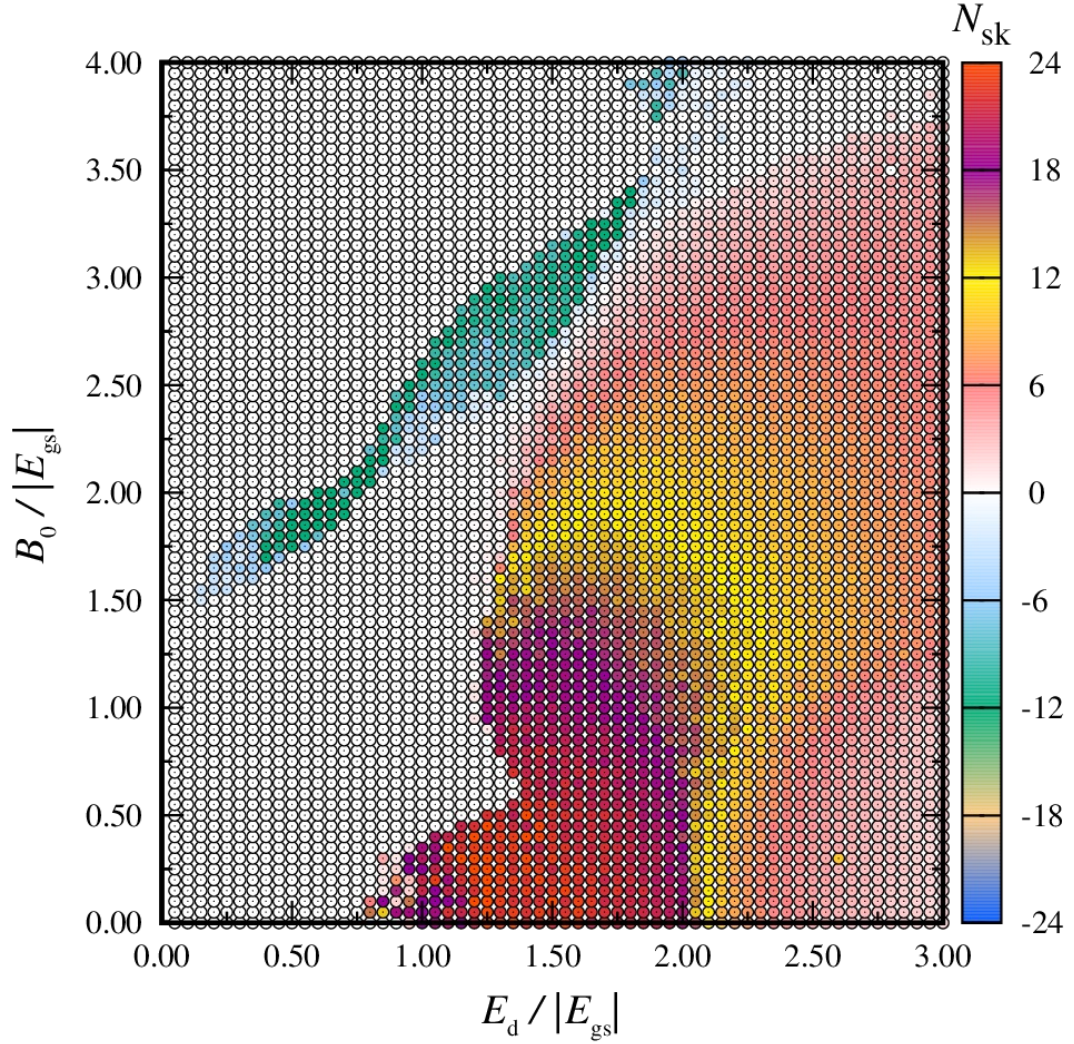} 
  \caption{\label{fig:PD1}
  Averaged skyrmion number $N_{\rm sk}$ as functions of $E_{\rm d}$ and $B_{0}$ at $\omega/|E_{\rm gs}|=7.5$ in the NESSs under the RCP electromagnetic field.
  }
  \end{center}
\end{figure}

First, we study the NESSs obtained by irradiating the circularly polarized electromagnetic field without the static magnetic field, namely, $B_{\mathrm{s}}^z=0$. 
As an initial spin configuration, we adopt the single-$Q$ spiral state, which corresponds to the ground state at $E_0= B_0 = 0$; we choose the spiral plane as the $xy$ plane, as the qualitatively similar results are obtained for the other spiral planes.

Figure~\ref{fig:PD1} shows the averaged skyrmion number $N_{\mathrm{sk}}$ as functions of the strength of the RCP electric field $E_{\mathrm{d}}=\lambda E_0$ and the RCP magnetic field $B_0$ in the NESSs; the skyrmion number becomes nonzero in a wide range of electromagnetic-field parameters, indicating the emergence of topological spin textures, i.e., the SkXs.
Across the line $B_0/E_{\mathrm{d}}=1$, the sign of the skyrmion number tends to be reversed; $N_{\rm sk}>0$ for $B_0/E_{\mathrm{d}}<1$ and $N_{\rm sk}<0$ for $B_0/E_{\mathrm{d}}>1$. 
Meanwhile, the stability region of the SkXs is asymmetric in terms of $B_0/E_{\mathrm{d}}=1$, with the RCP electric field being more favorable than the RCP magnetic field. 
Indeed, the SkX phase is stabilized around $1 \leq E_{\mathrm{d}}/|E_{\rm gs}|\leq 1.6$ solely by the RCP electric field even at $B_0 = 0$, whereas no SkX phase appears only by the RCP magnetic field at $E_{\rm d}=0$.
These results indicate that the RCP electric field induces a SkX with a positive skyrmion number, whereas the addition of an RCP magnetic field reverses the sign, leading to a SkX with a negative skyrmion number.

\begin{figure}[t!]
  \begin{center}
  \includegraphics[width=1.0\hsize]{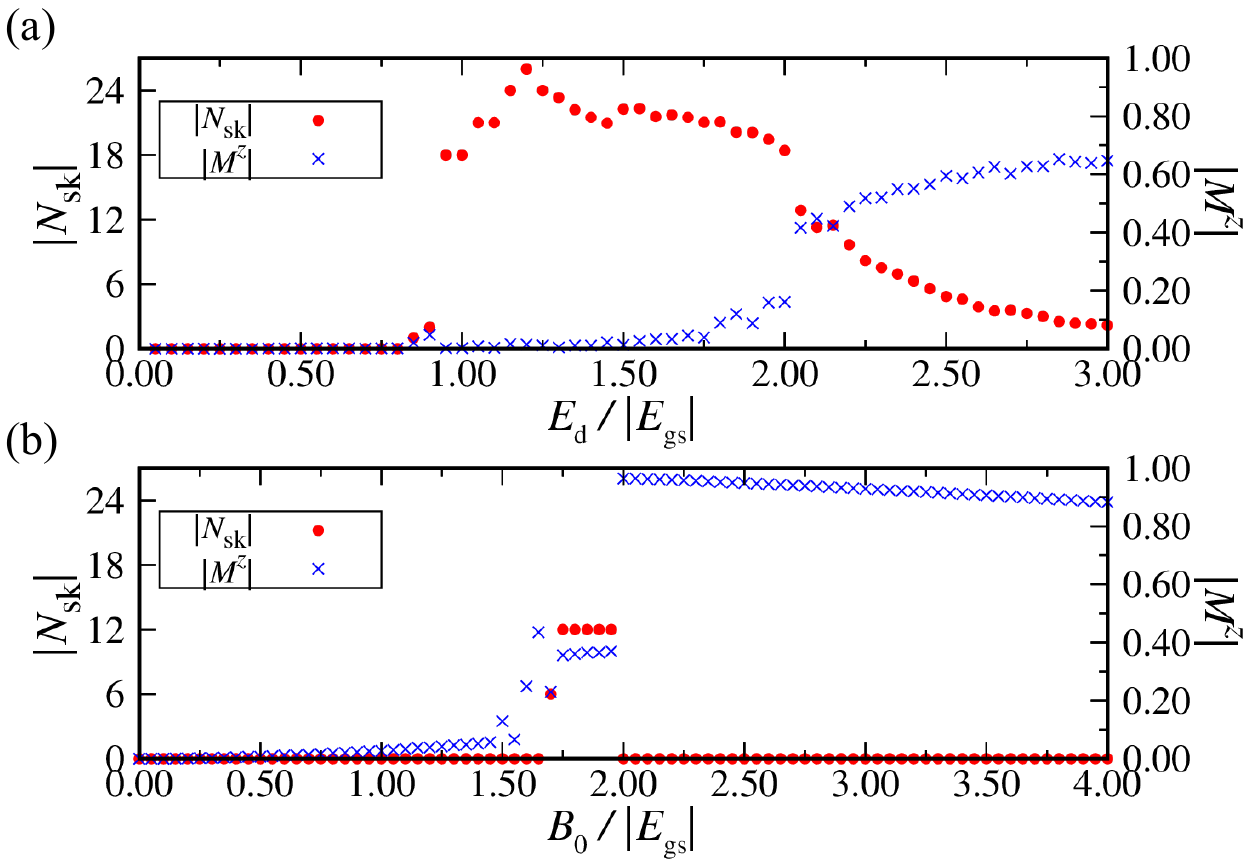} 
  \caption{\label{fig:Depend1}
  (a) $E_{\rm d}$ dependence of the absolute $N_{\rm sk}$ and averaged absolute out-of-plane magnetization $M^{z}$ at $B_0/E_{\mathrm{d}}=0.2$. 
  (b) $B_{0}$ dependence of the absolute $N_{\rm sk}$ and  the absolute $M^{z}$ at $B_0/E_{\mathrm{d}}=3.6$.
  }
  \end{center}
\end{figure}

\begin{figure*}[t!]
  \begin{center}
  \includegraphics[width=1.0\hsize]{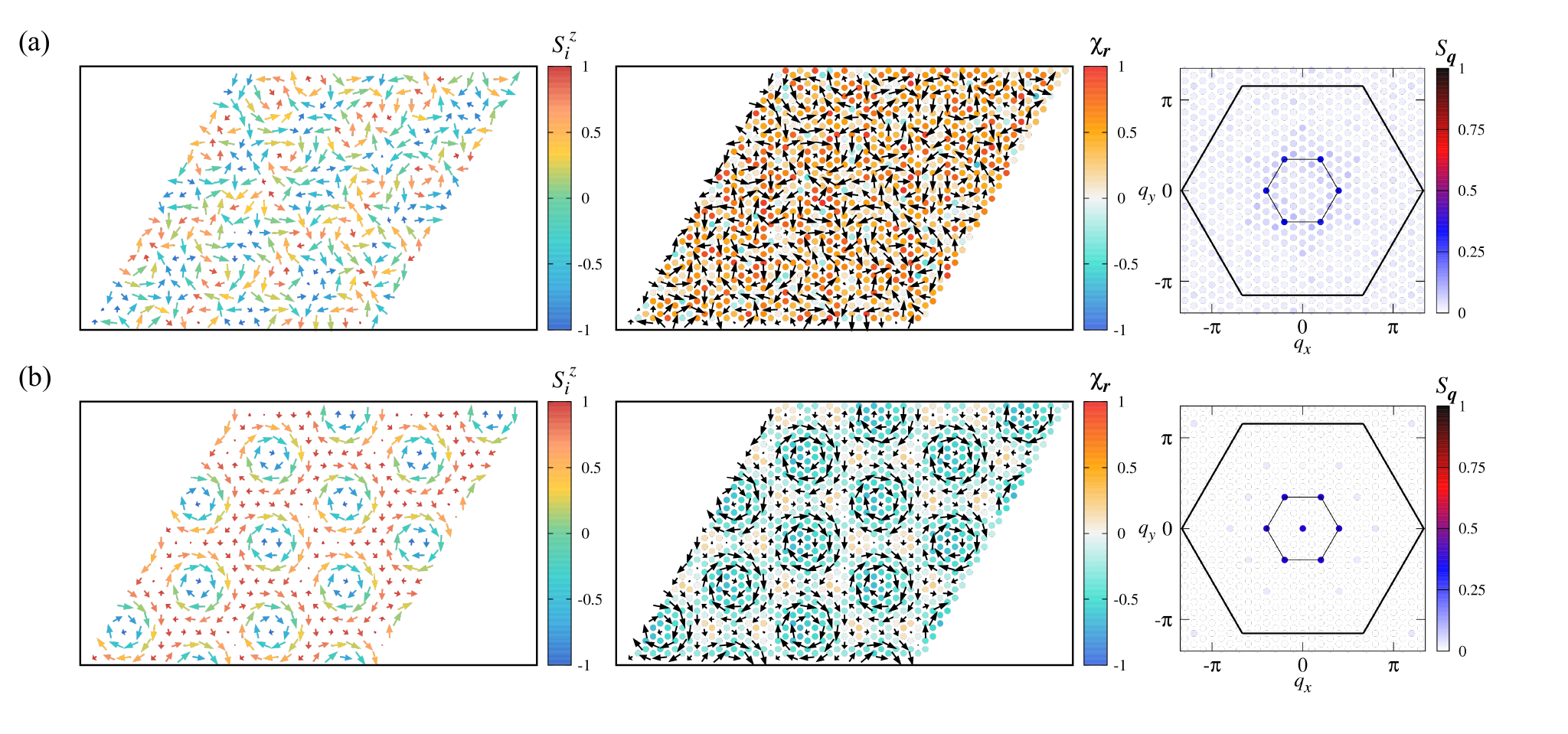}
  \caption{\label{fig:conf1}
  Snapshots of the NESSs under the RCP electric field with $\omega/|E_{\rm gs}|=7.5$;
  (a) ($E_{\rm d}/|E_{\rm gs}|, B_{0}/|E_{\rm gs}|)=(1.25, 0.25)$ and (b) ($E_{\rm d}/|E_{\rm gs}|,  B_{0}/|E_{\rm gs})=(0.5, 1.8)$.
  (Left panels) Real-space spin configurations; the arrows and color denote the spins $\bm{S}_i$ and the $z$ component $S^z_i$, respectively.
  (Middle panels) Local scalar spin chirality configurations; the arrows and color of the circles denote the spins $\bm{S}_i$ and the local scalar spin chirality $\chi_{\bm{r}}$, respectively.
  (Right panels) Reciprocal-space spins $S_{\bm{q}}$.
  The large hexagons represent the first Brillouin zone.
  The vertices of the small hexagons correspond to $\pm\boldsymbol{Q}^*_1$, $\pm\boldsymbol{Q}^*_2$, and $\pm\boldsymbol{Q}^*_3$.
  }
  \end{center}
\end{figure*}

The detailed dependence on $B_0$ and $E_{\mathrm{d}}$ at fixed $B_0/E_{\mathrm{d}}=0.2$ or $B_0/E_{\mathrm{d}}=3.6$ is presented in \Fig{fig:Depend1}.
When the electric-field effect is dominant compared to the magnetic-field one ($B_0 < E_{\mathrm{d}}$), the SkXs with $+18 \lesssim N_{\mathrm{sk}} \lesssim +24$ are induced, as shown in \Fig{fig:Depend1}(a).
The snapshot of the spin configuration of the SkX with $N_{\mathrm{sk}}=+24$ is shown in the left panel of \Fig{fig:conf1}(a).
This SkX consists of the skyrmion with $n_{\rm sk}= +2$~\cite{Yambe2024circular}; it has the positive scalar spin chirality and the superpositions of the ordering wave vectors $\bm{Q}_1^*$, $\bm{Q}_2^*$, and $\bm{Q}_3^*$, as shown in the middle and right panels of \Fig{fig:conf1}(a).
The sign of the scalar spin chirality is determined by the polarization of the light ($\delta$).  
Similarly, the SkX with $N_{\mathrm{sk}}=+18$ also corresponds to the periodic alignment of the skyrmions with $n_{\rm sk}= +2$, although the positions of the triple-$Q$ ordering wave vectors are located at $(\bm{Q}^{*}_1-\bm{Q}^{*}_3)/2$, $(\bm{Q}^{*}_2-\bm{Q}^{*}_1)/2$, and $(\bm{Q}^{*}_3-\bm{Q}^{*}_2)/2$.
Such SkXs with $n_{\rm sk}= +2$ have been reported even in thermodynamic systems by additionally considering the multi-spin interactions~\cite{Ozawa2017, Hayami_PhysRevB.99.094420, hayami2021phase, Eto_PhysRevB.104.104425} and magnetic anisotropic interactions~\cite{amoroso2020spontaneous, Hayami_PhysRevB.103.054422, Yambe2021skyrmion, Amoroso20212Dconductor, Wang_PhysRevB.103.104408, hayami2022multipleskyrmion} in addition to the frustrated exchange interactions, which indicates that the RCP electric field effectively plays a similar role to these effects, as described below~\cite{Yambe2024circular}.

In the other region for $E_{\mathrm{d}} > B_0$ in Fig.~\ref{fig:PD1}, the SkX with quantized skyrmion number is not clearly observed; for example, around $B_{0}/|E_{\rm gs}| \simeq 2.15$ and $E_{\rm d}/|E_{\rm gs}| \simeq 1.65$, the SkX with $N_{\mathrm{sk}} \sim +12$ is obitaned. 
We regard this state as a transient state of the melting of the topological spin textures from the SkXs with $N_{\mathrm{sk}}=+24$ and $N_{\mathrm{sk}}=+18$ to the topologically trivial state without $N_{\rm sk}$~\cite{Yambe2024circular}. 
Indeed, its spin structure factor exhibits a broadened peak structure rather than a sharp one shown in the right panel of Fig.~\ref{fig:conf1}.

On the other hand, the situtation where the magnetic-field effect is dominant ($B_0 > E_{\mathrm{d}}$) leads to the instability toward the another SkX with the skyrmion number $N_{\mathrm{sk}}=-12$, as shown in the case of $B_{0}/|E_{\rm gs}| \sim 1.8$ in \Fig{fig:Depend1}(b).
It consists of the Bloch-type skyrmion with $n_{\rm sk}= -1$ with a definite helicity and vorticity, as shown in the left panel of Fig.~\ref{fig:conf1}(b); it has the negative scalar spin chirality at the core and peaks at $\bm{q}=\bm{0}$ in addition to the ordering wave vectors $\bm{Q}_1^*$, $\bm{Q}_2^*$, and $\bm{Q}_3^*$, as shown in the middle and right panels of Fig.~\ref{fig:conf1}(b).
This SkX with $n_{\rm sk}= -1$ has been found in centrosymmetric magnets even without the dynamical electromagnetic fields, but other effects, such as the thermal fluctuations~\cite{Okubo2012, Mitsumoto_PhysRevB.105.094427} and single-ion anisotropy~\cite{Leonov2015, SZLin2016PhysRevB_GL, Hayami2016bubble, Hayami2021inplane}, are additionally taken into account.
In other words, the dynamical electromagnetic fields play a similar role in these effects.

To qualitatively understand the emergence of the SkXs with $n_{\rm sk}=+2$ and $n_{\rm sk}=-1$ in the phase diagram,
we introduce an effective static spin Hamiltonian for the model Hamiltonian in \Eq{H} within the Floquet formalism~\cite{Eckardt2015Floquet,Sato2016PRL,Yambe2023,Sato2025floquet}.
The effect of the time-periodic field is perturbatively incorporated into a time-independent Floquet Hamiltonian up to the lowest order of $\omega^{-1}$, which is given by~\cite{Yambe2023}
\begin{align}
  \mathcal{H}_{\mathrm{F}}
  &=\mathcal{H}_{\mathrm{static}}
    +\bm{\mathcal{B}}\cdot\sum_i\bm{S}_i+\mathcal{D}\sum_i\sum_{\alpha=1,2,3}\bm{e}_{\alpha}\cdot\left(\bm{S}_i\times \bm{S}_{i+\bm{e}_{\alpha}}\right)\notag\\
    &+\mathcal{T}\sum_{i}\sum_{\sigma=\pm 1}S_i^z\left(\bm{S}_{j+\sigma\bm{e}_1}\times \bm{S}_{j+\sigma\bm{e}_2}\right.\notag\\
    &+\bm{S}_{j+\sigma\bm{e}_2}\times \bm{S}_{j+\sigma\bm{e}_3}
    +\left.\bm{S}_{j+\sigma\bm{e}_3}\times \bm{S}_{j+\sigma\bm{e}_1}\right)^z\notag\\
    &+\mathcal{T}\sum_{\bigtriangleup,\bigtriangledown}\bm{S}_i\cdot\left(\bm{S}_j\times \bm{S}_k\right),\label{Floquet}
\end{align}
with
\begin{align}
  \bm{\mathcal{B}}&=\frac{\delta  B_0^2}{2\omega} \bm{e}_{z},\label{coeff1}\\
  \mathcal{D}&=\frac{\delta  E_0B_0\lambda}{2\omega},\label{coeff2}\\
  \intertext{and}
  \mathcal{T}&=-\frac{\sqrt{3}\delta E_0^2\lambda^2}{4\omega}.\label{coeff3}
\end{align}
The first term $\mathcal{H}_{\rm static}$ consists of the spin Hamiltonian $\mathcal{H}_{0}$ and the Zeeman coupling to the static magnetic field, and the other terms arise from the dynamical electromagnetic fields.
The second term in \Eq{Floquet} is the effective Zeeman coupling to the circularly polarized magnetic field proportional to $B_0^2$;
$\bm{e}_{z}$ in \Eq{coeff1} is the unit vector in the perpendicular direction to the $xy$ plane.
The third term in \Eq{Floquet} represents the effective field-induced chiral DM interaction arising from the coupling between electric and magnetic fields~\cite{Sato2016PRL};
$\bm{e}_1=(1, 0)$, $\bm{e}_2=(1/2, \sqrt{3}/2)$, and $\bm{e}_3=(-1/2, \sqrt{3}/2)$ are the vectors of the nearest-neighbor bonds.
The last two terms represent effective three-spin interactions induced by the second order of the electric field $E_0^2$~\cite{Yambe2024circular}, where the summation in the last term runs over both upward and downward unit triangles and the sites $(i,j,k)$ are labeled in a counterclockwise order. 

The effective interactions appearing in \Eq{Floquet} provide the reason why the SkXs with $n_{\rm sk}=+2$ and $n_{\rm sk}=-1$ are stabilized under dynamical electromagnetic fields intuitively. 
The SkX with $n_{\rm sk}=+2$, which emerges even in the absence of $B_0$, as shown in \Fig{fig:Depend1}(a), is attributed to the effective three-spin interactions that originate from the time-dependent electric field~\cite{Yambe2024circular}. 
This is understood from the fact that the coupling form in the three-spin interaction terms naturally favors noncoplanar spin textures with a finite scalar spin chirality.
Since the effective three-spin interaction depends on the light polarization $\delta$, the favorable sign of the scalar spin chirality is determined by $\delta$; the RCP with $\delta=+1$ (LCP with $\delta=-1$) selects the state with the positive (negative) scalar spin chirality. 
This is why the SkX with $n_{\rm sk}=+2$ is stabilized in the region for $B_0 < E_{\mathrm{d}}$. 

Meanwhile, the emergence of the SkXs with $n_{\rm sk}=-1$ stabilized in the region for $B_0 > E_{\mathrm{d}}$ in \Fig{fig:Depend1}(b) is qualitatively understood from the interplay between the effective Zeeman coupling and the effective chiral DM interactions, which becomes the origin of the magnetic-field-induced SkX, as found in static systems in noncentrosymmetric environments~\cite{Robler2006, Tokura2021, Hayami2024MaterToday}.
Thus, the circularly polarized electromagnetic fields mimic the situation in noncentrosymmetric chiral magnets when the magnetic-field effect is relatively dominant.
Furthermore, the Bloch-type helicity around the skyrmion core is naturally understood from the direction of the DM vector, which is parallel to the nearest-neighbor bond direction.

\subsection{Circularly polarized electric fields and static magnetic fields} \label{ssec:staticB}

\begin{figure}[t!]
  \begin{center}
  \includegraphics[width=1.0\hsize]{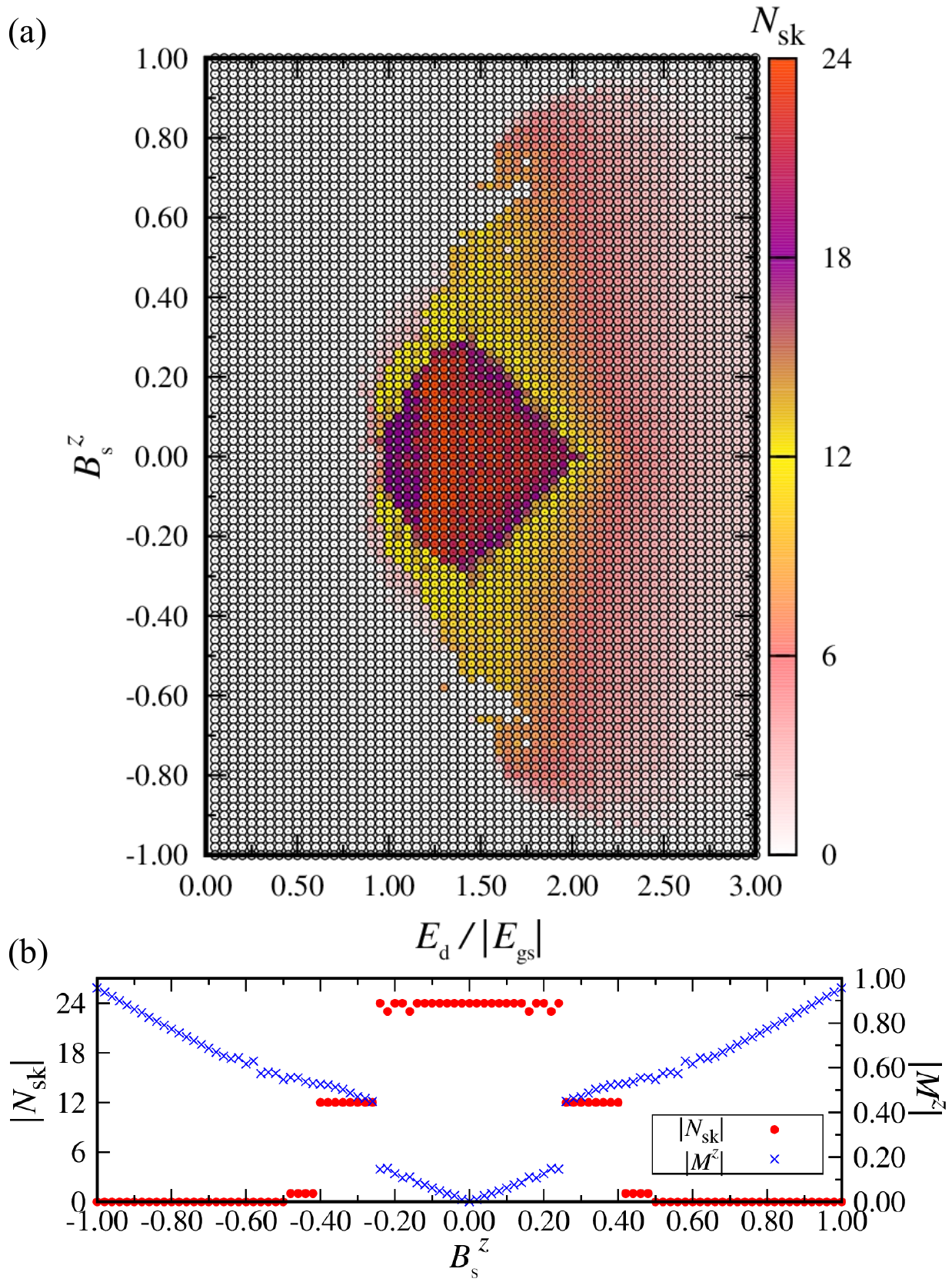} 
  \caption{\label{fig:PD2}
  (a) Averaged skyrmion number $N_{\rm sk}$ as functions of $E_{\rm d}$ and $B_{\rm s}^z$ at $\omega/|E_{\rm gs}|=7.5$ in the NESSs under the RCP electric field and the static magnetic field.
  (b) $B_{\mathrm{s}}^z$ dependence of the absolute $N_{\rm sk}$ and  the absolute $M^{z}$ at $E_{\mathrm{d}}/|E_{\rm gs}|=1.25$.
  }
  \end{center}
\end{figure}

\begin{figure*}[t!]
  \begin{center}
  \includegraphics[width=1.0\hsize]{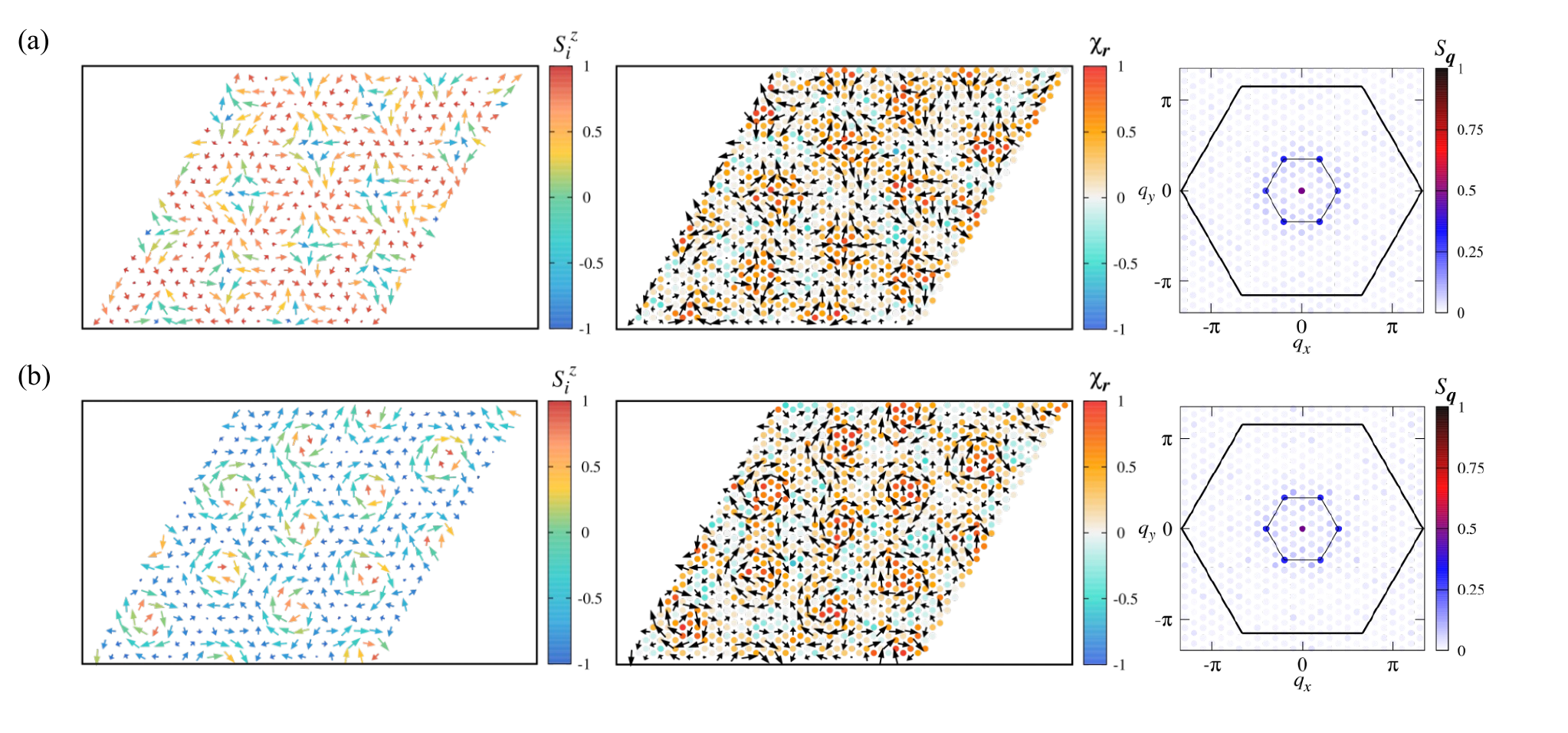} 
  \caption{\label{fig:conf2}
  Snapshots of the NESSs under the RCP electric field with $\omega/|E_{\rm gs}|=7.5$;
  (a) ($E_{\rm d}/|E_{\rm gs}|, B_{\mathrm{s}}^z)=(1.25, 0.3)$ and (b) ($E_{\rm d}/|E_{\rm gs}|, B_{\mathrm{s}}^z)=(1.25, -0.3)$.
  (Left panels) Real-space spin configurations; the arrows and color denote the spins $\bm{S}_i$ and the $z$ component $S^z_i$, respectively.
  (Middle panels) Local scalar spin chirality configurations; the arrows and color of the circles denote the spins $\bm{S}_i$ and the local scalar spin chirality $\chi_{\bm{r}}$, respectively.
  (Right panels) Reciprocal-space spins $S_{\bm{q}}$.
  The large hexagons represent the first Brillouin zone.
  The vertices of the small hexagons correspond to $\pm\boldsymbol{Q}^*_1$, $\pm\boldsymbol{Q}^*_2$, and $\pm\boldsymbol{Q}^*_3$.
  }
  \end{center}
\end{figure*}

Next, we study the NESSs by considering the effect of the static magnetic field rather than the circularly polarized magnetic field, namely, $B_0=0$. 
In other words, we consider the situation where the effect of the circularly polarized magnetic field is ignored by supposing $B_0 \ll E_{\mathrm{d}}$.
Similar to the previous case in \Sec{ssec:dynamicB}, we adopt the same initial spin configuration represented as the single-$Q$ spiral state.

Figure~\ref{fig:PD2}(a) shows the averaged skyrmion number $N_{\mathrm{sk}}$ as functions of the strength of the RCP electric field $E_{\mathrm{d}}$ and the static magnetic field $B_{\mathrm{s}}^z$ in the NESSs.
Induced topological spin textures appearing in the region around $1 \lesssim E_{\mathrm{d}}/|E_{\rm gs}| \lesssim 1.6$ uniformly exhibit the positive sign of the skyrmion number irrespective of the presence or absence of $B_{\mathrm{s}}^z$ and the magnetic-field directions, i.e., $N_{\mathrm{sk}}>0$.
Thus, different from \Sec{ssec:dynamicB}, the SkX with the negative skyrmion number is not stabilized under the RCP electric field and the static magnetic field. 
It is noted that the sign of the skyrmion number is reversed by the polarization of the light, as shown in Appendix~\ref{sec:appendixA}.

The $B_{\mathrm{s}}^z$ dependence of the skyrmion number and the magnetization at $E_{\mathrm{d}}=1.25$ is shown in \Fig{fig:PD2}(b).
For $|B_{\mathrm{s}}^z| \lesssim 0.24$, the SkXs with $N_{\mathrm{sk}}=+24$, which consist of the periodic array of the skyrmions with $n_{\rm sk}=+2$, are stabilized, similar to \Sec{ssec:dynamicB}.
By increasing $|B_{\mathrm{s}}^z|$, the skyrmion number decreases from $+24$ to $+12$.
The left panels of Fig.~\ref{fig:conf2} show real-space spin configurations of the SkXs with $N_{\mathrm{sk}}=+12$, which correspond to the periodic alignment of the skyrmions with $n_{\rm sk}=+1$ at $|B_{\mathrm{s}}^z| = 0.3$. 
The signs of the scalar spin chirality and the spins in reciprocal space shown in the middle and right panels of Fig.~\ref{fig:conf2}, in addition to the skyrmion numbers, are the same irrespective of the magnetic-field directions.
However, the real-space spin configuration seems to be different, as compared in the left panels of Figs.~\ref{fig:conf2}(a) and \ref{fig:conf2}(b). 
The skyrmion core is identified as the region with $S_i^z=-1$ and is characterized by the vorticity $-1$ for $B_{\mathrm{s}}^z>0$, as shown in Fig.~\ref{fig:conf2}(a). 
On the other hand, the skyrmion core is located at the region with $S_i^z=+1$ and is characterized by the vorticity $+1$ for $B_{\mathrm{s}}^z<0$, as shown in Fig.~\ref{fig:conf2}(b). 
Since both polarization and vorticity take opposite signs for the two situations, both spin configurations are characterized by the same sign of the skyrmion number and the scalar spin chirality, as shown in the middle panel of Figs.~\ref{fig:conf2}(a) and \ref{fig:conf2}(b). 
With a further increase of the magnetic field $|B_{\mathrm{s}}^z|$, the SkXs turn into the topologically trivial states without $N_{\rm sk}$.

We discuss the microscopic origin of the SkXs under the static magnetic field based on the Floquet Hamiltonian in Eqs.~(\ref{Floquet})--(\ref{coeff3}).
In contrast to the circularly polarized magnetic field, only the fourth and fifth terms are present; the effective chiral DM interaction is absent owing to $B_0=0$.
When the static magnetic field is small, the SkX with $N_{\rm sk}=+24$ is stabilized by the effective three-spin interactions, similar to the case in \Sec{ssec:dynamicB}. 
Meanwhile, the emergence of the SkX with $N_{\rm sk}=+12$ is attributed to the interplay between the effective three-spin interactions and the Zeeman coupling to the static magnetic field; the former tends to favor a triple-$Q$ spin configuration and the latter tends to favor a superposition of spiral waves rather than the sinusoidal waves~\cite{Yambe2021skyrmion}.
Without two-spin in-plane magnetic anisotropy, the helicity and vorticity of the skyrmion core are not uniquely determined in this situation. 
In addition, the favorable sign of the skyrmion number is determined by the light polarization.
Thus, the sign of the skyrmion number in the magnetic-field-induced SkX is affected by the light polarization, with the LCP electric field, rather than the RCP one, leading to the SkXs with the negative skyrmion number.

\begin{figure}[t!]
  \begin{center}
  \includegraphics[width=1.0\hsize]{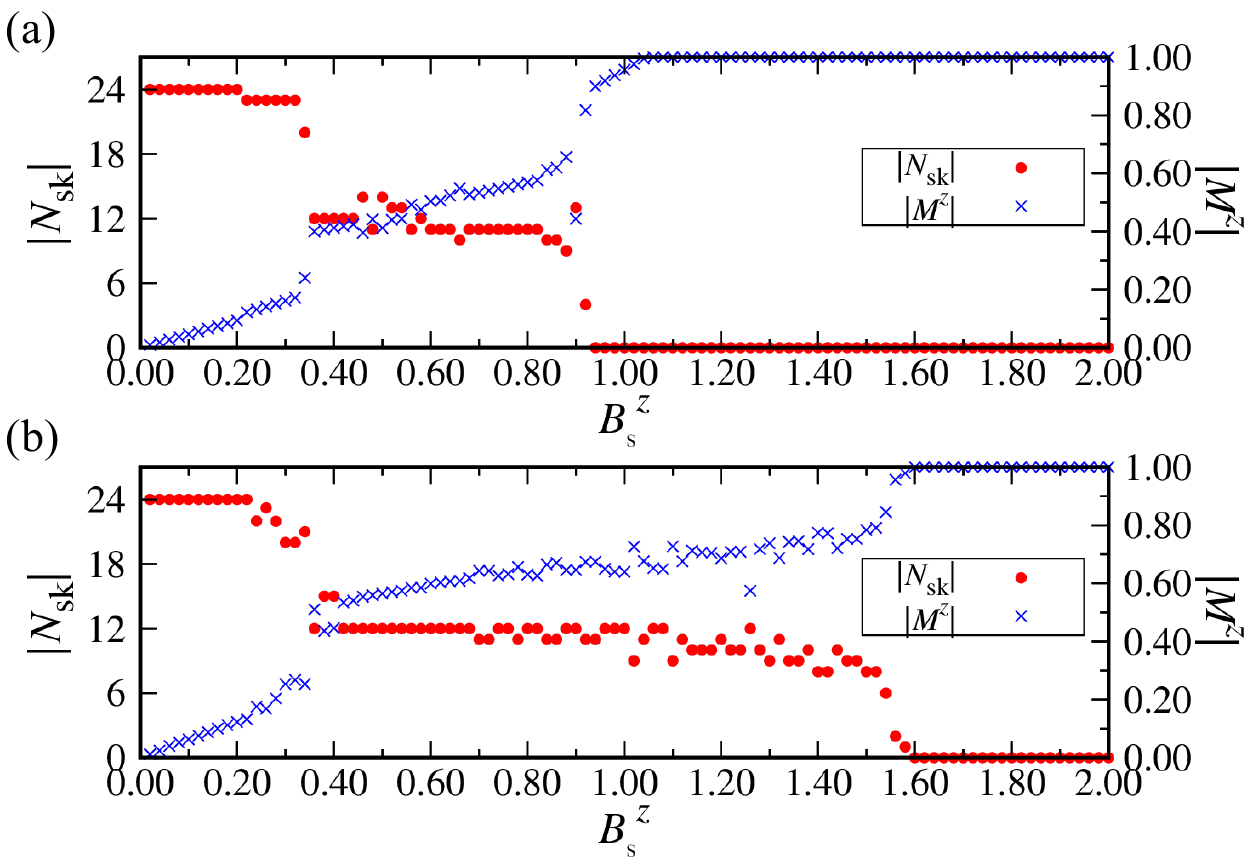} 
  \caption{\label{fig:Depend2}
  $B_{\mathrm{s}}^z$ dependence of the absolute $N_{\rm sk}$ and averaged absolute $M^{z}$ under the static magnetic field at $E_{\mathrm{d}}/|E_{\rm gs}|=1.25$. 
  The results in (a) and (b) stand for absence and presence of the RCP electric field with $\omega/|E_{\rm gs}|=7.5$ after the system reaches the NESS  without the magnetic field, respectively. 
  }
  \end{center}
\end{figure}

The stability region of the magnetic-field-induced SkX with $N_{\rm sk}=+12$ is largely affected by the timing of the application of the RCP electric field and the static magnetic field. 
To demonstrate this, we perform the following setups. 
First, we obtain the NESS under the RCP electric field at $E_{\mathrm{d}}/|E_{\rm gs}|=1.25$ by adopting the single-$Q$ spiral configuration with the spiral plane in the $xy$ plane as the initial one. 
Subsequently, a static magnetic field is applied to the NESS, and two situations are examined: (i) the RCP electric field is turned off, and (ii) the RCP electric field remains applied.
The former and latter results as a function of $B_{\rm s}^{z}$ are summarized in Figs.~\ref{fig:Depend2}(a) and \ref{fig:Depend2}(b), respectively. 
The simulations are independently performed for different $B^z_{\rm s}$.

As shown in Figs.~\ref{fig:Depend2}(a) and \ref{fig:Depend2}(b), both results show the topological phase transition between the SkXs with $N_{\rm sk}=+24$ and $N_{\rm sk}=+12$ at $B_{\mathrm{s}}^z \simeq 0.36$.
Meanwhile, the stability region of the SkX with $N_{\rm sk}=+12$ is substantially extended when the RCP electric field is kept. 
This tendency can be attributed to the effective three-spin interactions induced by the RCP electric field, which enhance the stability of spin configurations with finite scalar spin chirality, such as the SkX. 
However, the critical magnetic field of the transition between the SkXs with $N_{\rm sk}=+24$ and $N_{\rm sk}=+12$ is not affected by the presence or absence of the RCP electric field, which implies that the higher-order SkX possessing large magnetization $|M^z|$ becomes energetically unstable, resulting in the phase transition to another SkX.

Moreover, compared to \Fig{fig:PD2}(b), where both the RCP electric field and the static magnetic field are simultaneously applied at $t=0$, both the SkXs with $N_{\rm sk}=+24$ and $N_{\rm sk}=+12$ are more stabilized in the present setup.
In the previous setup, the topological phase transition between the SkXs with $N_{\rm sk}=+24$ and $N_{\rm sk}=+12$ ($N_{\rm sk}=+12$ and $N_{\rm sk}\sim0$) occurs at $B_{\mathrm{s}}^z \simeq 0.26$ $(B_{\mathrm{s}}^z \simeq 0.42)$, whereas the larger magnetic fields are required to obtain the same topological phase transitions as shown in Fig~\ref{fig:Depend2}.
This features the robustness of topological spin textures against the external fields once they are generated.
Thus, the timing of the application of the RCP electric field and the static magnetic field is important in obtaining the SkXs as a NESS.

\section{SUMMARY} \label{sec:summary}
\begin{table}[h!]
\centering
\caption{
The classification of the skyrmion number within the magnetic unit cell $n_{\mathrm{sk}}$ under time-dependent electromagnetic fields and the static magnetic field. 
We omit the SkX with $n_{\mathrm{sk}}=\pm 2$ under the circularly polarized or static magnetic fields.
\label{table:nsk}
}
\begin{tabular}{c|cc}
\hline\hline
 \diagbox{Magnetic field}{Electric field} & RCP $(\delta=+1)$ & LCP $(\delta=-1)$ \\
\hline
No magnetic fields & $+2$ & $-2$ \\
\hline
RCP $(\delta=+1)$ & $-1$ & --  \\
LCP $(\delta=-1)$ &  -- & $ +1$ \\
\hline
$B_{\mathrm{s}}^z>0$ & $+1$ & $ -1$\\
$B_{\mathrm{s}}^z<0$ & $+1$ & $ -1$\\
\hline\hline
\end{tabular}
\end{table}

We have theoretically investigated the stability of SkXs with different skyrmion numbers and their phase transition in frustrated magnets irradiating with time-dependent electromagnetic fields. 
Our analysis is based on the calculations for the classical Heisenberg model on the two-dimensional triangular lattice, where the effect of the electric field is incorporated through the inverse DM mechanism.
We have numerically solved the LLG equation by considering two setups on external magnetic fields in addition to the circularly polarized electric field: One is the circularly polarized magnetic field and the other is the static magnetic field.

Starting the simulations from the single-$Q$ spiral spin configuration, we have elucidated a variety of the SkXs and their topological transitions.
In both cases, the SkX with $n_{\rm sk}=+2$ is stabilized under the RCP electric field, when the effect of the magnetic field is small. 
Meanwhile, a qualitatively difference is found in the magnetic-field-induced SkXs; the RCP magnetic field induces the SkX with $n_{\rm sk}=-1$ and the static magnetic field induces the SkX with $n_{\rm sk}=+1$. 
The appearance of the magnetic-field-induced SkX in the former case is accounted for by the interplay between the effective chiral DM interactions and dynamical Zeeman coupling, while that in the latter case is accountd for by the interplay between the effective three-spin interactions and static Zeeman coupling, by deriving the effective time-independent spin model based on the Floquet formalism. 
Since the sign of the effective interactions are reversed by the reversal of the light polarization, the SkXs with the opposite signs of the skyrmion number are obtained when the LCP electromagnetic field is applied. 
We summarize the relationship between the sign of the skyrmion number in the SkX and the applied electromagnetic fields in Table~\ref{table:nsk}. 
We have also demonstrated that the stability region of the SkXs is largely extended by controlling the timing of the applied circularly polarized electric field and static magnetic field. 
Our findings extend the framework for generating and controlling SkXs through time-dependent electromagnetic fields in collaboration with the static magnetic field, offering new insights into both the fundamental understanding and practical manipulation of topological spin structures. 

\begin{acknowledgments}
This research was supported by JSPS KAKENHI Grants Numbers JP22H00101, JP22H01183, JP23H04869, JP23K03288, JP23K20827, and by JST CREST (JPMJCR23O4) and JST FOREST (JPMJFR2366). 
\end{acknowledgments}

\appendix
\section{Case of the left circularly polarized electric field}\label{sec:appendixA}

\begin{figure*}[t!]
  \begin{center}
  \includegraphics[width=1.0\hsize]{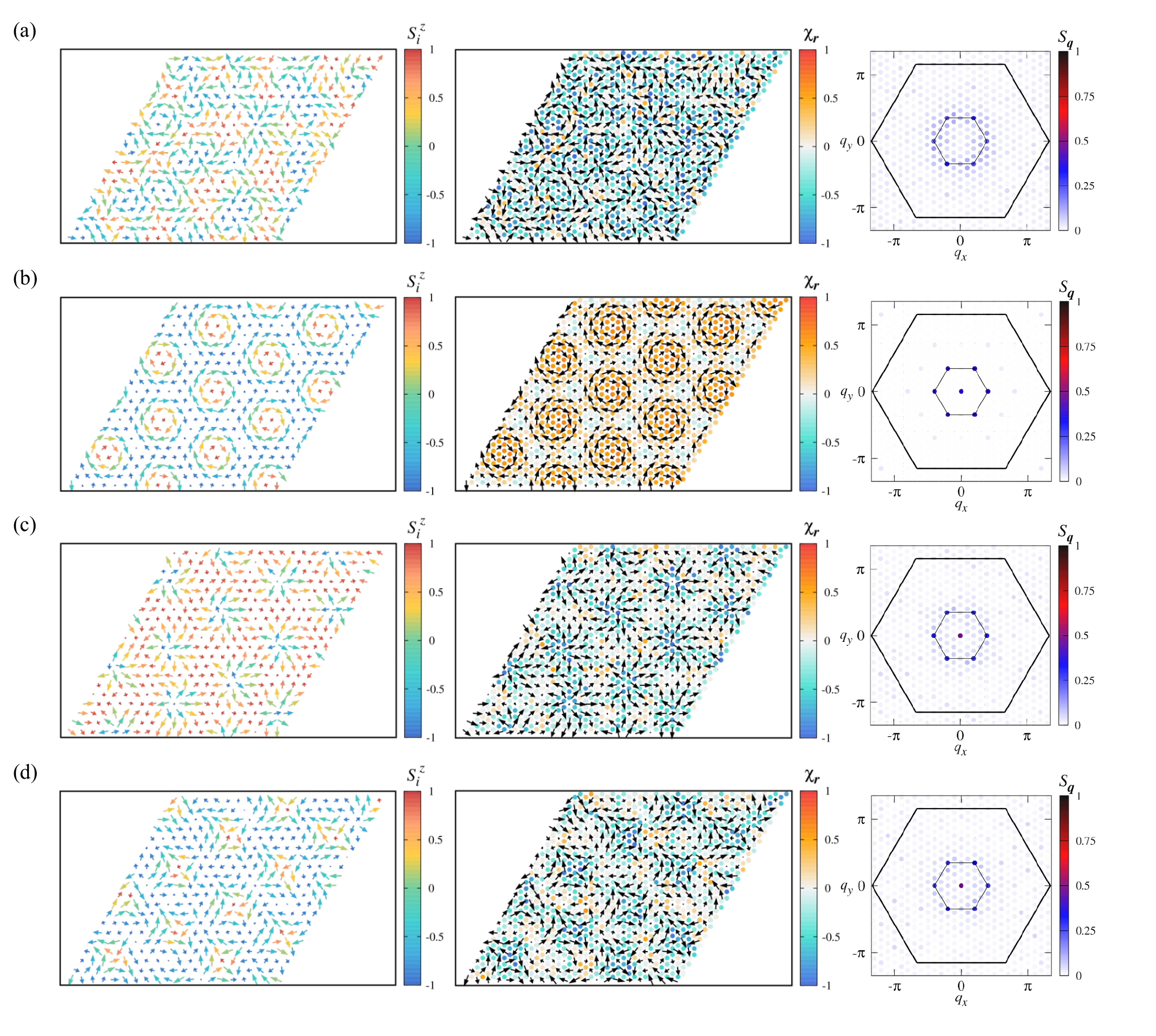} 
  \caption{\label{fig:conf3}
  Snapshots of the NESSs under the LCP electric field with $\omega/|E_{\rm gs}|=7.5$;
  (a) ($E_{\rm d}/|E_{\rm gs}|, B_0/|E_{\rm gs}|, B_{\mathrm{s}}^z)=(1.25, 0.25, 0)$, 
  (b) ($E_{\rm d}/|E_{\rm gs}|, B_0/|E_{\rm gs}|, B_{\mathrm{s}}^z)=(1.25, 1.8, 0)$,
  (c) ($E_{\rm d}/|E_{\rm gs}|, B_0/|E_{\rm gs}|, B_{\mathrm{s}}^z)=(1.25, 0, 0.3)$, and 
  (d) ($E_{\rm d}/|E_{\rm gs}|, B_0/|E_{\rm gs}|, B_{\mathrm{s}}^z)=(1.25, 0, -0.3)$.
  (Left panels) Real-space spin configurations; the arrows and color denote the spins $\bm{S}_i$ and the $z$ component $S^z_i$, respectively.
  (Middle panels) Local scalar spin chirality configurations; the arrows and color of the circles denote the spins $\bm{S}_i$ and the local scalar spin chirality $\chi_{\bm{r}}$, respectively.
  (Right panels) Reciprocal-space spins $S_{\bm{q}}$.
  The large hexagons represent the first Brillouin zone.
  The vertices of the small hexagons correspond to $\pm\boldsymbol{Q}^*_1$, $\pm\boldsymbol{Q}^*_2$, and $\pm\boldsymbol{Q}^*_3$.
  }
  \end{center}
\end{figure*}

In Sec.~\ref{sec:results}, we present the results under the RCP electric field. 
Since the LCP electric field in combination with the magnetic fields leads to the same magnitude of the effective interaction but with a different sign, the stability region of the SkXs is not altered.
Meanwhile, the sign of the skyrmion number is reversed. 
The results for representative parameters are shown in Fig.~\ref{fig:conf3}, where Figs.~\ref{fig:conf3}(a,b) [Figs.~\ref{fig:conf3}(c,d)] present the case under the LCP electromagnetic field (the combination of the LCP electric field and the static magnetic field).  
In each case, the SkXs with the opposite skyrmion number to those obtained under the RCP fields are generated.

\bibliographystyle{apsrev4-2}
\bibliography{PRB_full}

@article{PhysRevResearch.6.013228,
  title = {Electric-field control of magnetic anisotropies: Applications to Kitaev spin liquids and topological spin textures},
  author = {Furuya, Shunsuke C. and Sato, Masahiro},
  journal = {Phys. Rev. Res.},
  volume = {6},
  issue = {1},
  pages = {013228},
  numpages = {19},
  year = {2024},
  month = {Mar},
  publisher = {American Physical Society},
  doi = {10.1103/PhysRevResearch.6.013228},
  url = {https://link.aps.org/doi/10.1103/PhysRevResearch.6.013228}
}

@article{Mitsumoto_PhysRevB.105.094427,
  title = {Skyrmion crystal in the {RKKY} system on the two-dimensional triangular lattice},
  author = {Mitsumoto, Kota and Kawamura, Hikaru},
  journal = {Phys. Rev. B},
  volume = {105},
  issue = {9},
  pages = {094427},
  numpages = {13},
  year = {2022},
  month = {Mar},
  publisher = {American Physical Society},
  doi = {10.1103/PhysRevB.105.094427},
  url = {https://link.aps.org/doi/10.1103/PhysRevB.105.094427}
}

@article{Wang_PhysRevB.103.104408,
  title = {Meron, skyrmion, and vortex crystals in centrosymmetric tetragonal magnets},
  author = {Wang, Zhentao and Su, Ying and Lin, Shi-Zeng and Batista, Cristian D.},
  journal = {Phys. Rev. B},
  volume = {103},
  issue = {10},
  pages = {104408},
  numpages = {12},
  year = {2021},
  month = {Mar},
  publisher = {American Physical Society},
  doi = {10.1103/PhysRevB.103.104408},
  url = {https://link.aps.org/doi/10.1103/PhysRevB.103.104408}
}

@article{Eto_PhysRevB.104.104425,
  title = {Dynamical switching of magnetic topology in microwave-driven itinerant magnet},
  author = {Eto, Rintaro and Mochizuki, Masahito},
  journal = {Phys. Rev. B},
  volume = {104},
  issue = {10},
  pages = {104425},
  numpages = {13},
  year = {2021},
  month = {Sep},
  publisher = {American Physical Society},
  doi = {10.1103/PhysRevB.104.104425},
  url = {https://link.aps.org/doi/10.1103/PhysRevB.104.104425}
}

@article{hayami2021phase,
  title={Phase shift in skyrmion crystals},
  author={Hayami, Satoru and Okubo, Tsuyoshi and Motome, Yukitoshi},
  journal={Nat. Commun.},
  volume={12},
  pages={6927},
  year={2021},
  doi={10.1038/s41467-021-27083-0},
}

@article{Adams_PhysRevLett.107.217206,
  title = {{Long-Range Crystalline Nature of the Skyrmion Lattice in {MnSi}}},
  author = {Adams, T. and M\"uhlbauer, S. and Pfleiderer, C. and Jonietz, F. and Bauer, A. and Neubauer, A. and Georgii, R. and B\"oni, P. and Keiderling, U. and Everschor, K. and Garst, M. and Rosch, A.},
  journal = {Phys. Rev. Lett.},
  volume = {107},
  issue = {21},
  pages = {217206},
  numpages = {5},
  year = {2011},
  month = {Nov},
  publisher = {American Physical Society},
  doi = {10.1103/PhysRevLett.107.217206},
  url = {https://link.aps.org/doi/10.1103/PhysRevLett.107.217206}
}

@inproceedings{adams2010skyrmion,
  title={Skyrmion lattice domains in {Fe$_{1- x}$Co$_x$Si}},
  author={Adams, T and M{\"u}hlbauer, S and Neubauer, A and M{\"u}nzer, W and Jonietz, F and Georgii, R and Pedersen, B and B{\"o}ni, P and Rosch, A and Pfleiderer, C},
  booktitle={J. Phys.: Conf. Ser.},
  volume={200},
  number={3},
  pages={032001},
  year={2010},
  organization={IOP Publishing},
  doi={10.1088/1742-6596/200/3/032001}
}

@article{Shindou_PhysRevLett.87.116801,
  title = {{Orbital Ferromagnetism and Anomalous {Hall} Effect in Antiferromagnets on the Distorted fcc Lattice}},
  author = {Shindou, Ryuichi and Nagaosa, Naoto},
  journal = {Phys. Rev. Lett.},
  volume = {87},
  issue = {11},
  pages = {116801},
  numpages = {4},
  year = {2001},
  month = {Aug},
  doi = {10.1103/PhysRevLett.87.116801},
  publisher = {American Physical Society}
}

@article{Ye_PhysRevLett.83.3737,
  title = {Berry Phase Theory of the Anomalous {Hall} Effect: Application to Colossal Magnetoresistance Manganites},
  author = {Ye, Jinwu and Kim, Yong Baek and Millis, A. J. and Shraiman, B. I. and Majumdar, P. and Te\ifmmode \check{s}\else \v{s}\fi{}anovi\ifmmode \acute{c}\else \'{c}\fi{}, Z.},
  journal = {Phys. Rev. Lett.},
  volume = {83},
  issue = {18},
  pages = {3737--3740},
  numpages = {0},
  year = {1999},
  month = {Nov},
  publisher = {American Physical Society},
  doi = {10.1103/PhysRevLett.83.3737},
  url = {https://link.aps.org/doi/10.1103/PhysRevLett.83.3737}
}

@article{Loss_PhysRevB.45.13544,
  title = {Persistent currents from {Berry's} phase in mesoscopic systems},
  author = {Loss, Daniel and Goldbart, Paul M.},
  journal = {Phys. Rev. B},
  volume = {45},
  issue = {23},
  pages = {13544--13561},
  numpages = {0},
  year = {1992},
  month = {Jun},
  publisher = {American Physical Society},
  doi = {10.1103/PhysRevB.45.13544},
}

@article{Bogdanov89,
  title = {Thermodynamically stable ``vortices" in magnetically ordered crystals: The mixed state of magnets},
  author = {Bogdanov, A. N. and Yablonskii, D. A.},
  journal = {Sov. Phys. JETP},
  volume = {68},
  pages = {101},
  year = {1989},
}

@article{RK1954PhysRev,
  title = {Indirect Exchange Coupling of Nuclear Magnetic Moments by Conduction Electrons},
  author = {Ruderman, M. A. and Kittel, C.},
  journal = {Phys. Rev.},
  volume = {96},
  issue = {1},
  pages = {99--102},
  numpages = {0},
  year = {1954},
  month = {Oct},
  publisher = {American Physical Society},
  doi = {10.1103/PhysRev.96.99},
  url = {https://link.aps.org/doi/10.1103/PhysRev.96.99}
}

@article{Kasuya1956PTP,
    author = {Kasuya, Tadao},
    title = {A Theory of Metallic Ferro- and Antiferromagnetism on Zener's Model},
    journal = {Progr. Theor. Phys.},
    volume = {16},
    number = {1},
    pages = {45-57},
    year = {1956},
    month = {07},
    issn = {0033-068X},
    doi = {10.1143/PTP.16.45},
    url = {https://doi.org/10.1143/PTP.16.45},
}

@article{Yosida1957PhysRev,
  title = {Magnetic Properties of Cu-Mn Alloys},
  author = {Yosida, Kei},
  journal = {Phys. Rev.},
  volume = {106},
  issue = {5},
  pages = {893--898},
  numpages = {0},
  year = {1957},
  month = {Jun},
  publisher = {American Physical Society},
  doi = {10.1103/PhysRev.106.893},
  url = {https://link.aps.org/doi/10.1103/PhysRev.106.893}
}

@article{KT1973JPhysC,
doi = {10.1088/0022-3719/6/7/010},
url = {https://doi.org/10.1088/0022-3719/6/7/010},
year = {1973},
month = {apr},
publisher = {},
volume = {6},
number = {7},
pages = {1181},
author = {J M Kosterlitz and D J Thouless},
title = {Ordering, metastability and phase transitions in two-dimensional systems},
journal = {J. Phys. C},
}

@article{Kosterlitz1974JPhysC,
doi = {10.1088/0022-3719/7/6/005},
url = {https://doi.org/10.1088/0022-3719/7/6/005},
year = {1974},
month = {mar},
publisher = {},
volume = {7},
number = {6},
pages = {1046},
author = {J M Kosterlitz},
title = {The critical properties of the two-dimensional xy model},
journal = {J. Phys. C},
}

@article{Berry1984,
  title = {Quantal phase factors accompanying adiabatic changes},
  author = {Michael Victor Berry},
  journal = {Proc. R. Soc. Lond. A},
  volume = {392},
  issue = {1802},
  pages = {45--47},
  year = {1984},
  month = {Mar},
  publisher = {Royal Society},
  doi = {10.1098/rspa.1984.0023},
  url = {https://doi.org/10.1098/rspa.1984.0023}
}

@article{TKNN1982PRL,
  title = {Quantized Hall Conductance in a Two-Dimensional Periodic Potential},
  author = {Thouless, D. J. and Kohmoto, M. and Nightingale, M. P. and den Nijs, M.},
  journal = {Phys. Rev. Lett.},
  volume = {49},
  issue = {6},
  pages = {405--408},
  numpages = {0},
  year = {1982},
  month = {Aug},
  publisher = {American Physical Society},
  doi = {10.1103/PhysRevLett.49.405},
  url = {https://link.aps.org/doi/10.1103/PhysRevLett.49.405}
}

@article{KOHMOTO1985Ann.Phys,
title = {Topological invariant and the quantization of the Hall conductance},
journal = {Ann. Phys. (N.Y.)},
volume = {160},
number = {2},
pages = {343-354},
year = {1985},
issn = {0003-4916},
doi = {10.1016/0003-4916(85)90148-4},
url = {https://www.sciencedirect.com/science/article/pii/0003491685901484},
author = {Mahito Kohmoto},
}

@article{Tanaka2019Neural,
title = {Recent advances in physical reservoir computing: A review},
journal = {Neural Netw.},
volume = {115},
pages = {100-123},
year = {2019},
issn = {0893-6080},
doi = {10.1016/j.neunet.2019.03.005},
url = {https://www.sciencedirect.com/science/article/pii/S0893608019300784},
author = {Gouhei Tanaka and Toshiyuki Yamane and Jean Benoit Héroux and Ryosho Nakane and Naoki Kanazawa and Seiji Takeda and Hidetoshi Numata and Daiju Nakano and Akira Hirose},
}

@article{Yokouchi2022Sciadv,
author = {Tomoyuki Yokouchi and Satoshi Sugimoto and Bivas Rana  and Shinichiro Seki  and Naoki Ogawa  and Yuki Shiomi and Shinya Kasai and Yoshichika Otani },
title = {Pattern recognition with neuromorphic computing using magnetic field-induced dynamics of skyrmions},
journal = {Sci. Adv.},
volume = {8},
number = {39},
pages = {eabq5652},
year = {2022},
doi = {10.1126/sciadv.abq5652},
URL = {https://www.science.org/doi/abs/10.1126/sciadv.abq5652}
}

@article{Lee2024NatMater,
author={Lee, Oscar and Wei, Tianyi and Stenning, Kilian D. and Gartside, Jack C. and Prestwood, Dan and Seki, Shinichiro and Aqeel, Aisha and Karube, Kosuke and Kanazawa, Naoya and Taguchi, Yasujiro and Back, Christian and Tokura, Yoshinori and Branford, Will R. and Kurebayashi, Hidekazu},
title={Task-adaptive physical reservoir computing},
journal={Nat. Mater.},
year={2024},
month={Jan},
day={01},
volume={23},
number={1},
pages={79-87},
issn={1476-4660},
doi={10.1038/s41563-023-01698-8},
url={https://doi.org/10.1038/s41563-023-01698-8}
}

@article{Sato2025floquet,
  title={Floquet theory and applications in open quantum and classical systems},
  author = {Masahiro Sato and Tatsuhiko N. Ikeda},
  journal = {arXiv:2508.01783},
}

@article{Sato2016PRL,
  title = {Laser-Driven Multiferroics and Ultrafast Spin Current Generation},
  author = {Sato, Masahiro and Takayoshi, Shintaro and Oka, Takashi},
  journal = {Phys. Rev. Lett.},
  volume = {117},
  issue = {14},
  pages = {147202},
  numpages = {5},
  year = {2016},
  month = {Sep},
  publisher = {American Physical Society},
  doi = {10.1103/PhysRevLett.117.147202},
  url = {https://link.aps.org/doi/10.1103/PhysRevLett.117.147202}
}

@article{Eckardt2015Floquet,
  title = {High-frequency approximation for periodically driven quantum systems from a Floquet-space perspective},
  author = {André Eckardt and Egidijus Anisimovas},
  journal = {New J. Phys.},
  volume = {17},
  pages = {093039},
  year = {2015},
  month = {Sep},
  doi = {10.1088/1367-2630/17/9/093039},
}

@article{Shirato_JPSJ.94.063601_bimeron,
author = {Shirato ,Tatsuya and Yambe ,Ryota and Hayami ,Satoru},
title = {Bimeron Crystals by a Linearly Polarized AC Electric Field in Frustrated Magnets},
journal = {J. Phys. Soc. Jpn.},
volume = {94},
number = {6},
pages = {063601},
year = {2025},
doi = {10.7566/JPSJ.94.063601},
URL = {https://doi.org/10.7566/JPSJ.94.063601},
}

@article{rackauckas2017differentialequations,
  title={Differentialequations.jl--a performant and feature-rich ecosystem for solving differential equations in julia},
  author={Rackauckas, Christopher and Nie, Qing},
  journal={J. Open Res. Softw.},
  volume={5},
  number={1},
  pages={15},
  year={2017},
  publisher={Ubiquity Press}
}

@article{Hayami_PhysRevB.99.094420,
  title = {Effect of magnetic anisotropy on skyrmions with a high topological number in itinerant magnets},
  author = {Hayami, Satoru and Motome, Yukitoshi},
  journal = {Phys. Rev. B},
  volume = {99},
  issue = {9},
  pages = {094420},
  numpages = {8},
  year = {2019},
  month = {Mar},
  publisher = {American Physical Society},
  doi = {10.1103/PhysRevB.99.094420},
  url = {https://link.aps.org/doi/10.1103/PhysRevB.99.094420}
}

@article{Hayami_PhysRevB.103.054422,
  title = {Noncoplanar multiple-{$Q$} spin textures by itinerant frustration: Effects of single-ion anisotropy and bond-dependent anisotropy},
  author = {Hayami, Satoru and Motome, Yukitoshi},
  journal = {Phys. Rev. B},
  volume = {103},
  issue = {5},
  pages = {054422},
  numpages = {36},
  year = {2021},
  month = {Feb},
  publisher = {American Physical Society},
  doi = {10.1103/PhysRevB.103.054422},
  url = {https://link.aps.org/doi/10.1103/PhysRevB.103.054422}
}

@article{Yi_PhysRevB.80.054416,
  title = {Skyrmions and anomalous {Hall} effect in a {Dzyaloshinskii}-{Moriya} spiral magnet},
  author = {Yi, Su Do and Onoda, Shigeki and Nagaosa, Naoto and Han, Jung Hoon},
  journal = {Phys. Rev. B},
  volume = {80},
  issue = {5},
  pages = {054416},
  numpages = {6},
  year = {2009},
  month = {Aug},
  publisher = {American Physical Society},
  doi = {10.1103/PhysRevB.80.054416},
  url = {http://link.aps.org/doi/10.1103/PhysRevB.80.054416}
}

@article{Yambe2021skyrmion,
  title={Skyrmion crystals in centrosymmetric itinerant magnets without horizontal mirror plane},
  author={Yambe, Ryota and Hayami, Satoru},
  journal={Sci. Rep.},
  volume={11},
  pages={11184},
  year={2021},
  doi={10.1038/s41598-021-90308-1},
  publisher={Nature Publishing Group}
}

@article{amoroso2020spontaneous,
  title={Spontaneous skyrmionic lattice from anisotropic symmetric exchange in a {Ni}-halide monolayer},
  author={Amoroso, Danila and Barone, Paolo and Picozzi, Silvia},
  journal={Nat. Commun.},
  volume={11},
  number={1},
  pages={5784},
  year={2020},
  doi={10.1038/s41467-020-19535-w},
  publisher={Nature Publishing Group}
}

@article{skyrme1962unified,
  title={A unified field theory of mesons and baryons},
  author={Skyrme, Tony Hilton Royle},
  journal={Nucl. Phys.},
  volume={31},
  pages={556--569},
  year={1962},
  publisher={Elsevier}
}

@article{yu2010real,
  title={Real-space observation of a two-dimensional skyrmion crystal},
  author={Yu, X Z and Onose, Y and Kanazawa, N and Park, J H and Han, J H and Matsui, Y and Nagaosa, N and Tokura, Y},
  journal={Nature},
  volume={465},
  number={7300},
  pages={901--904},
  year={2010},
  doi = {10.1038/nature09124},
  publisher={Nature Publishing Group}
}

@article{Hirschberger_10.1088/1367-2630/abdef9,
	author={Max Hirschberger and Satoru Hayami and Yoshinori Tokura},
	title={Nanometric skyrmion lattice from anisotropic exchange interactions in a centrosymmetric host},
	journal={New J. Phys.},
	volume = {23},
	pages = {023039},
	year={2021},
	doi={10.1088/1367-2630/abdef9},
	url={http://iopscience.iop.org/article/10.1088/1367-2630/abdef9},
}

@article{hirschberger2019skyrmion,
  title={Skyrmion phase and competing magnetic orders on a breathing kagome lattice},
  author={Hirschberger, Max and Nakajima, Taro and Gao, Shang and Peng, Licong and Kikkawa, Akiko and Kurumaji, Takashi and Kriener, Markus and Yamasaki, Yuichi and Sagayama, Hajime and Nakao, Hironori and Ohishi, Kazuki and Kakurai, Kazuhisa and Taguchi, Yasujiro and Yu, Xiuzhen and Arima,Taka-hisa and Tokura, Yoshinori},
  journal={Nat. Commun.},
  volume={10},
  number={1},
  pages={5831},
  year={2019},
  publisher={Nature Publishing Group},
  doi = {10.1038/s41467-019-13675-4},
}

@article{SZLin2016PhysRevB_GL,
  title = {Ginzburg-Landau theory for skyrmions in inversion-symmetric magnets with competing interactions},
  author = {Lin, Shi-Zeng and Hayami, Satoru},
  journal = {Phys. Rev. B},
  volume = {93},
  issue = {6},
  pages = {064430},
  numpages = {16},
  year = {2016},
  month = {Feb},
  publisher = {American Physical Society},
  doi = {10.1103/PhysRevB.93.064430},
  url = {https://link.aps.org/doi/10.1103/PhysRevB.93.064430}
}

@article{hayami2022multipleskyrmion,
  title={Multiple skyrmion crystal phases by itinerant frustration in centrosymmetric tetragonal magnets},
  author={Hayami, Satoru},
  journal={J. Phys. Soc. Jpn.},
  volume={91},
  number={2},
  pages={023705},
  year={2022},
  publisher={The Physical Society of Japan},
  doi={10.7566/JPSJ.91.023705},
}

@article{Okubo2012,
   author = {Tsuyoshi Okubo and Sungki Chung and Hikaru Kawamura},
   doi = {10.1103/PhysRevLett.108.017206},
   issn = {00319007},
   issue = {1},
   journal = {Phys. Rev. Lett.},
   month = {1},
   pages = {017206},
   title = {Multiple-q states and the Skyrmion lattice of the triangular-lattice Heisenberg antiferromagnet under magnetic fields},
   volume = {108},
   year = {2012},
}

@article{Yambe2024circular,
   author = {Ryota Yambe and Satoru Hayami},
   doi = {10.1103/PHYSREVB.110.014428/FIGURES/8/THUMBNAIL},
   issn = {24699969},
   issue = {1},
   journal = {Phys. Rev. B},
   month = {7},
   pages = {014428},
   publisher = {American Physical Society},
   title = {Dynamical generation of skyrmion and bimeron crystals by a circularly polarized electric field in frustrated magnets},
   volume = {110},
   year = {2024},
}

@article{Yambe2023,
   author = {Ryota Yambe and Satoru Hayami},
   doi = {10.1103/PhysRevB.108.064420},
   issn = {24699969},
   issue = {6},
   journal = {Phys. Rev. B},
   month = {8},
   pages = {064420},
   publisher = {American Physical Society},
   title = {Symmetry analysis of light-induced magnetic interactions via Floquet engineering},
   volume = {108},
   year = {2023},
}

@article{Katsura2005,
   author = {Hosho Katsura and Naoto Nagaosa and Alexander V. Balatsky},
   doi = {10.1103/PhysRevLett.95.057205},
   issn = {00319007},
   issue = {5},
   journal = {Phys. Rev. Lett.},
   month = {7},
   pages = {057205},
   title = {Spin current and magnetoelectric effect in noncollinear magnets},
   volume = {95},
   year = {2005},
}

@article{Sergienko2006,
   author = {I. A. Sergienko and E. Dagotto},
   doi = {10.1103/PhysRevB.73.094434},
   issn = {10980121},
   issue = {9},
   journal = {Phys. Rev. B},
   pages = {094434},
   title = {Role of the Dzyaloshinskii-Moriya interaction in multiferroic perovskites},
   volume = {73},
   year = {2006},
}

@article{Mostovoy2006,
   author = {Maxim Mostovoy},
   doi = {10.1103/PhysRevLett.96.067601},
   issn = {10797114},
   issue = {6},
   journal = {Phys. Rev. Lett.},
   pages = {067601},
   publisher = {American Physical Society},
   title = {Ferroelectricity in spiral magnets},
   volume = {96},
   year = {2006},
}

@article{Zhang2020skyrmion,
  author = {Xichao Zhang and Yan Zhou and Kyung Mee Song and Tae-Eon Park and Jing Xia and Motohiko Ezawa and Xiaoxi Liu and Weisheng Zhao and Guoping Zhao and Seonghoon Woo},
  title = {Skyrmion-electronics: writing, deleting, reading and processing magnetic skyrmions toward spintronic applications},
  journal = {J. Phys.: Condens. Matter},
  volume = {32},
  number = {14},
  pages = {143001},
  year = {2020},
  doi = {10.1088/1361-648X/ab5488},
  publisher = {IOP Publishing}
}

@article{Hayami2021itinerant,
   author = {Satoru Hayami and Yukitoshi Motome},
   doi = {10.1088/1361-648X/AC1A30},
   issn = {0953-8984},
   issue = {44},
   journal = {J. Phys.: Condens. Matter},
   month = {8},
   pages = {443001},
   pmid = {34343975},
   publisher = {IOP Publishing},
   title = {Topological spin crystals by itinerant frustration},
   volume = {33},
   url = {https://iopscience.iop.org/article/10.1088/1361-648X/ac1a30 https://iopscience.iop.org/article/10.1088/1361-648X/ac1a30/meta},
   year = {2021},
}

@article{Ohgushi2000,
   author = {K. Ohgushi and S. Murakami and N. Nagaosa},
   doi = {10.1103/PhysRevB.62.R6065},
   issn = {01631829},
   issue = {10},
   journal = {Phys. Rev. B},
   month = {9},
   pages = {R6065},
   title = {Spin anisotropy and quantum Hall effect in the <i>kagomé</i> lattice: Chiral spin state based on a ferromagnet},
   volume = {62},
   year = {2000},
}

@article{Kurumaji2019,
   author = {Takashi Kurumaji and Taro Nakajima and Max Hirschberger and Akiko Kikkawa and Yuichi Yamasaki and Hajime Sagayama and Hironori Nakao and Yasujiro Taguchi and Taka hisa Arima and Yoshinori Tokura},
   doi = {10.1126/SCIENCE.AAU0968/SUPPL_FILE/AAU0968_KURUMAJI_SM.PDF},
   issn = {10959203},
   issue = {6456},
   journal = {Science},
   month = {8},
   pages = {914-918},
   pmid = {31395744},
   publisher = {American Association for the Advancement of Science},
   title = {Skyrmion lattice with a giant topological Hall effect in a frustrated triangular-lattice magnet},
   volume = {365},
   url = {https://www.science.org/doi/10.1126/science.aau0968},
   year = {2019},
}

@article{Bogdanov1994,
   author = {A. Bogdanov and A. Hubert},
   doi = {10.1016/0304-8853(94)90046-9},
   issn = {0304-8853},
   issue = {3},
   journal = {J. Magn. Magn. Mater.},
   month = {12},
   pages = {255-269},
   publisher = {North-Holland},
   title = {Thermodynamically stable magnetic vortex states in magnetic crystals},
   volume = {138},
   year = {1994},
}

@article{Muhlbauer2009,
   author = {S. Mühlbauer and B. Binz and F. Jonietz and C. Pfleiderer and A. Rosch and A. Neubauer and R. Georgii and P. Böni},
   doi = {10.1126/SCIENCE.1166767/SUPPL_FILE/MUEHLBAUER.SOM.PDF},
   issn = {00368075},
   issue = {5916},
   journal = {Science},
   month = {2},
   pages = {915-919},
   pmid = {19213914},
   publisher = {American Association for the Advancement of Science},
   title = {Skyrmion lattice in a chiral magnet},
   volume = {323},
   url = {https://www.science.org/doi/10.1126/science.1166767},
   year = {2009},
}

@article{Neubauer2009,
   author = {A. Neubauer and C. Pfleiderer and B. Binz and A. Rosch and R. Ritz and P. G. Niklowitz and P. Böni},
   doi = {10.1103/PHYSREVLETT.102.186602/FIGURES/4/THUMBNAIL},
   issn = {10797114},
   issue = {18},
   journal = {Phys. Rev. Lett.},
   month = {5},
   pages = {186602},
   pmid = {19518895},
   publisher = {American Physical Society},
   title = {Topological hall effect in the a phase of MnSi},
   volume = {102},
   year = {2009},
}

@article{Dzyaloshinsky1958,
   author = {I. Dzyaloshinsky},
   doi = {10.1016/0022-3697(58)90076-3},
   issn = {0022-3697},
   issue = {4},
   journal = {J. Phys. Chem. Solids},
   month = {1},
   pages = {241-255},
   publisher = {Pergamon},
   title = {A thermodynamic theory of “weak” ferromagnetism of antiferromagnetics},
   volume = {4},
   year = {1958},
}

@article{Moriya1960,
   author = {Tôru Moriya},
   doi = {10.1103/PhysRev.120.91},
   issn = {0031899X},
   issue = {1},
   journal = {Phys. Rev.},
   pages = {91-98},
   title = {Anisotropic Superexchange Interaction and Weak Ferromagnetism},
   volume = {120},
   year = {1960},
}

@article{Robler2006,
   author = {U. K. Rößler and A. N. Bogdanov and C. Pfleiderer},
   doi = {10.1038/nature05056},
   issn = {1476-4687},
   issue = {7104},
   journal = {Nature},
   month = {8},
   pages = {797-801},
   publisher = {Nature Publishing Group},
   title = {Spontaneous skyrmion ground states in magnetic metals},
   volume = {442},
   url = {https://www.nature.com/articles/nature05056},
   year = {2006},
}

@article{Ozawa2017,
   author = {Ryo Ozawa and Satoru Hayami and Yukitoshi Motome},
   doi = {10.1103/PHYSREVLETT.118.147205/FIGURES/2/THUMBNAIL},
   issn = {10797114},
   issue = {14},
   journal = {Phys. Rev. Lett.},
   month = {4},
   pages = {147205},
   pmid = {28430467},
   publisher = {American Physical Society},
   title = {Zero-Field Skyrmions with a High Topological Number in Itinerant Magnets},
   volume = {118},
   year = {2017},
}

@article{Koshibae2014,
   author = {Wataru Koshibae and Naoto Nagaosa},
   doi = {10.1038/ncomms6148},
   issn = {2041-1723},
   issue = {1},
   journal = {Nat. Commun.},
   month = {10},
   pages = {1-11},
   publisher = {Nature Publishing Group},
   title = {Creation of skyrmions and antiskyrmions by local heating},
   volume = {5},
   url = {https://www.nature.com/articles/ncomms6148},
   year = {2014},
}

@article{Mochizuki2015,
   author = {Masahito Mochizuki and Yoshio Watanabe},
   doi = {10.1063/1.4929727/30781},
   issn = {00036951},
   issue = {8},
   journal = {Appl. Phys. Lett.},
   month = {8},
   pages = {082409},
   publisher = {American Institute of Physics Inc.},
   title = {Writing a skyrmion on multiferroic materials},
   volume = {107},
   url = {/aip/apl/article/107/8/082409/30781/Writing-a-skyrmion-on-multiferroic-materials},
   year = {2015},
}

@article{Mochizuki2016,
   author = {Masahito Mochizuki},
   doi = {10.1002/AELM.201500180},
   issn = {2199-160X},
   issue = {1},
   journal = {Adv. Electron. Mater.},
   month = {1},
   pages = {1500180},
   publisher = {John Wiley & Sons, Ltd},
   title = {Creation of Skyrmions by Electric Field on Chiral-Lattice Magnetic Insulators},
   volume = {2},
   url = {https://onlinelibrary.wiley.com/doi/full/10.1002/aelm.201500180 https://onlinelibrary.wiley.com/doi/abs/10.1002/aelm.201500180 https://onlinelibrary.wiley.com/doi/10.1002/aelm.201500180},
   year = {2016},
}

@article{Huang2018,
   author = {Ping Huang and Marco Cantoni and Alex Kruchkov and Jayaraman Rajeswari and Arnaud Magrez and Fabrizio Carbone and Henrik M. Rønnow},
   doi = {10.1021/ACS.NANOLETT.8B02097/SUPPL_FILE/NL8B02097_SI_002.AVI},
   issn = {15306992},
   issue = {8},
   journal = {Nano Lett.},
   month = {8},
   pages = {5167-5171},
   pmid = {30040904},
   publisher = {American Chemical Society},
   title = {In Situ Electric Field Skyrmion Creation in Magnetoelectric Cu2OSeO3},
   volume = {18},
   url = {https://pubs.acs.org/doi/abs/10.1021/acs.nanolett.8b02097},
   year = {2018},
}

@article{Hayami2016bubble,
   author = {Satoru Hayami and Shi Zeng Lin and Cristian D. Batista},
   doi = {10.1103/PHYSREVB.93.184413/FIGURES/10/THUMBNAIL},
   issn = {24699969},
   issue = {18},
   journal = {Phys. Rev. B},
   month = {5},
   pages = {184413},
   publisher = {American Physical Society},
   title = {Bubble and skyrmion crystals in frustrated magnets with easy-axis anisotropy},
   volume = {93},
   year = {2016},
}

@article{Amoroso20212Dconductor,
   author = {Danila Amoroso and Paolo Barone and Silvia Picozzi},
   doi = {10.3390/NANO11081873},
   issn = {2079-4991},
   issue = {8},
   journal = {Nanomaterials},
   month = {7},
   pages = {1873},
   publisher = {Multidisciplinary Digital Publishing Institute},
   title = {Interplay between Single-Ion and Two-Ion Anisotropies in Frustrated 2D Semiconductors and Tuning of Magnetic Structures Topology},
   volume = {11},
   url = {https://www.mdpi.com/2079-4991/11/8/1873/htm https://www.mdpi.com/2079-4991/11/8/1873},
   year = {2021},
}

@article{Yu2012,
   author = {X. Z. Yu and N. Kanazawa and W. Z. Zhang and T. Nagai and T. Hara and K. Kimoto and Y. Matsui and Y. Onose and Y. Tokura},
   doi = {10.1038/ncomms1990},
   issn = {2041-1723},
   issue = {1},
   journal = {Nat. Commun.},
   month = {8},
   pages = {988},
   pmid = {22871807},
   publisher = {Nature Publishing Group},
   title = {Skyrmion flow near room temperature in an ultralow current density},
   volume = {3},
   url = {https://www.nature.com/articles/ncomms1990},
   year = {2012},
}

@article{Berg1981,
   author = {B. Berg and M. Lüscher},
   doi = {10.1016/0550-3213(81)90568-X},
   issn = {0550-3213},
   issue = {2},
   journal = {Nucl. Phys. B},
   month = {8},
   pages = {412-424},
   publisher = {North-Holland},
   title = {Definition and statistical distributions of a topological number in the lattice O(3) σ-model},
   volume = {190},
   year = {1981},
}

@article{Tokura2021,
   author = {Yoshinori Tokura and Naoya Kanazawa},
   doi = {10.1021/ACS.CHEMREV.0C00297/ASSET/IMAGES/LARGE/CR0C00297_0026.JPEG},
   issn = {15206890},
   issue = {5},
   journal = {Chem. Rev.},
   month = {3},
   pages = {2857-2897},
   pmid = {33164494},
   publisher = {American Chemical Society},
   title = {Magnetic Skyrmion Materials},
   volume = {121},
   url = {https://pubs.acs.org/doi/full/10.1021/acs.chemrev.0c00297},
   year = {2021},
}

@article{Nagaosa2013,
   author = {Naoto Nagaosa and Yoshinori Tokura},
   doi = {10.1038/nnano.2013.243},
   issn = {17483395},
   issue = {12},
   journal = {Nat. Nanotech.},
   month = {12},
   pages = {899-911},
   pmid = {24302027},
   publisher = {Nature Publishing Group},
   title = {Topological properties and dynamics of magnetic skyrmions},
   volume = {8},
   year = {2013},
}

@article{Bogdanov1989,
   author = {A N Bogdanov and D A Yablonskii},
   journal = {Sov. Phys. JETP},
   pages = {101},
   title = {Thermodynamically stable "vortices" in magnetically ordered crystals. The mixed state of magnets},
   volume = {68},
   year = {1989},
}

@article{Hayami2024MaterToday,
title = {Stabilization mechanisms of magnetic skyrmion crystal and multiple-Q states based on momentum-resolved spin interactions},
journal = {Mater. Today Quantum},
volume = {3},
pages = {100010},
year = {2024},
issn = {2950-2578},
doi = {10.1016/j.mtquan.2024.100010},
url = {https://www.sciencedirect.com/science/article/pii/S2950257824000106},
author = {Satoru Hayami and Ryota Yambe}
}

@article{Hayami2021inplane,
  title = {In-plane magnetic field-induced skyrmion crystal in frustrated magnets with easy-plane anisotropy},
  author = {Hayami, Satoru},
  journal = {Phys. Rev. B},
  volume = {103},
  issue = {22},
  pages = {224418},
  numpages = {8},
  year = {2021},
  month = {Jun},
  publisher = {American Physical Society},
  doi = {10.1103/PhysRevB.103.224418},
  url = {https://link.aps.org/doi/10.1103/PhysRevB.103.224418}
}

@article{Leonov2015,
   author = {A. O. Leonov and M. Mostovoy},
   doi = {10.1038/ncomms9275},
   issn = {2041-1723},
   issue = {1},
   journal = {Nat. Commun.},
   month = {9},
   pages = {8275},
   publisher = {Nature Publishing Group},
   title = {Multiply periodic states and isolated skyrmions in an anisotropic frustrated magnet},
   volume = {6},
   url = {https://www.nature.com/articles/ncomms9275},
   year = {2015}
}

@article{Tokura2014,
author = {Tokura, Yoshinori and Seki, Shinichiro and Nagaosa, Naoto},
doi = {10.1088/0034-4885/77/7/076501},
url = {https://dx.doi.org/10.1088/0034-4885/77/7/076501},
year = {2014},
month = {jul},
publisher = {IOP Publishing},
volume = {77},
number = {7},
pages = {076501},
title = {Multiferroics of spin origin},
journal = {Rep. Prog. Phys.},
}

\end{document}